\documentclass{PNAStwoModified}

\usepackage[dvipdf]{graphicx}
\usepackage{amssymb,amsfonts,amsmath}

\newcommand{\Np}{N_+}\newcommand{\Nm}{N_-}\newcommand{\np}{n_+}\newcommand{\nm}{n_-}
\newcommand{\ntilde}{\tilde{n}}\newcommand{\nptilde}{\tilde{n}_+}\newcommand{\nmtilde}{\tilde{n}_-}
\newcommand{\eps}{\epsilon}
\newcommand{\epsO}{\epsilon_0}\newcommand{\epsOP}{\epsilon_{0+}}\newcommand{\epsOM}{\epsilon_{0-}}\newcommand{\epsOPM}{\epsilon_{0\pm}}
\newcommand{\KM}{K_-}

\newcommand{\piO}{\pi_0}\newcommand{\piOP}{\pi_{0+}}\newcommand{\piOM}{\pi_{0-}}\newcommand{\piOPM}{\pi_{0\pm}}
\newcommand{\vP}{v_+}\newcommand{\vM}{v_-}
\newcommand{\vB}{v_B}\newcommand{\vBM}{v_{B-}}\newcommand{\vBPM}{v_{B\pm}}
\newcommand{\vF}{v_F}\newcommand{\vFP}{v_{F+}}\newcommand{\vFM}{v_{F-}}\newcommand{\vFPM}{v_{F\pm}}
\newcommand{\Fs}{F_s}\newcommand{\FsP}{F_{s+}}\newcommand{\FsM}{F_{s-}}
\newcommand{\Fd}{F_d}\newcommand{\FdP}{F_{d+}}\newcommand{\FdM}{F_{d-}}\newcommand{\FdPM}{F_{d\pm}}
\newcommand{\fM}{f_-}
\newcommand{\FC}{F_c}\newcommand{\vC}{v_c}\newcommand{\FsC}{F_{Sc}}\newcommand{\Fp}{F_+}\newcommand{\Fm}{F_-}\newcommand{\Fext}{F_{\rm ext}}

\newcommand{\mycite}[1]{\cite{#1}}\newcommand{\Eq}[1]{Eq.~\eqref{#1}}
\newcommand{\EqsAnd}[2]{Eqs.~\eqref{#1} and \eqref{#2}}\newcommand{\Eqss}[3]{Eqs.~\eqref{#1}, \eqref{#2}, \eqref{#3}}

\newcommand{\ie}{i.e. }\newcommand{\eg}{e.g. }\newcommand{\resp}{resp. }
\newcommand{\invitro}{\textit{in vitro}}\newcommand{\invivo}{\textit{in vivo}}
\newcommand{\Invivo}{\textit{In vivo}}
\newcommand{\Drosophila}{\textit{Drosophila}}

 \newcommand{\N}{(0)}\renewcommand{\P}{($+$)}\newcommand{\M}{($-$)}
 \newcommand{\PM}{($-+$)}\newcommand{\PN}{(0$+$)}\newcommand{\NM}{($-$0)}\newcommand{\PMN}{($-$0$+$)}

\newcommand{\WtII}{Wt II}\newcommand{\WtIII}{Wt III}
\newcommand{\DhcWeak}{$Dhc^{6-10}/+$}\newcommand{\DhcStrong}{$Dhc^{8-1}/+$}\newcommand{\DhcBoth}{$Dhc^{8-1}/Dhc^{6-10}$}

\usepackage[colorlinks=true,citecolor=black,linkcolor=black,bookmarksnumbered=true,hyperfootnotes=true]{hyperref}
\hypersetup{
pdfauthor={Melanie J.I. M{\"u}ller, Stefan Klumpp, Reinhard Lipowsky},
pdftitle={Tug-of-war as a cooperative mechanism for bidirectional cargo transport by molecular motors},
pdfsubject={},
pdfkeywords={bidirectional mocement, cytoskeletal motor, intracellular transport, stochastic processes}
}

\newcommand{\FigTugOfWarCartoon}{Fig.~1}
\newcommand{\FigCargoMotorFluct}{Fig.~2}
\newcommand{\FigMotilityCharSym}{Fig.~3}
\newcommand{\FigMotilityCharAsym}{Fig.~4}
\newcommand{\FigMotilityWtII}{Fig.~5}
\newcommand{\TabMotorParameters}{Table~1}
\newcommand{\FigMotDiagSym}{Fig.~6}
\newcommand{\FigMotDiagAsym}{Fig.~7}
\newcommand{\FigCorrVelRunLength}{Fig.~8}
\newcommand{\FigHistoCorrVelRunLength}{Fig.~9}
\newcommand{\FigDoubleExp}{Fig.~10}
\newcommand{\FigWtIIPauseTimes}{Fig.~11}
\newcommand{\TabFitPara}{Table~2}
\newcommand{\TabFitResults}{Table~3}
\newcommand{\TabFitResultsB}{Table~4}

\begin{document}


\title{\begin{minipage}{\textwidth}Tug-of-war as a cooperative mechanism for\newline bidirectional cargo transport by molecular motors\end{minipage}}

\author{%
Melanie J.I. M{\"u}ller\affil{1}{Max Planck Institute of Colloids and Interfaces, Science Park Golm, 14424 Potsdam, Germany}, %
Stefan Klumpp\affil{2}{Center for Theoretical Biological Physics, University of California San Diego, La Jolla, CA 92093-0374, USA}, \and %
Reinhard Lipowsky\affil{1}{}
\thanks{To whom correspondence should be addressed. E-mail: lipowsky@mpikg.mpg.de}\affil{1}{}%
}

\maketitle


\begin{article}

\begin{abstract}

Intracellular transport is based on molecular motors that pull cargos along cytoskeletal filaments. One motor species always moves in one direction, \eg conventional kinesin moves to the microtubule plus end, while cytoplasmic dynein moves to the microtubule minus end. However, many cellular cargos are observed to move bidirectionally, involving both plus-end and minus-end directed motors. The presumably simplest mechanism for such bidirectional transport is provided by a tug-of-war between the two motor species. This mechanism is studied theoretically using the load-dependent transport properties of individual motors as measured in single-molecule experiments. In contrast to previous expectations, such a tug-of-war is found to be highly cooperative and to exhibit seven different motility regimes depending on the precise values of the single motor parameters. The sensitivity of the transport process to small parameter changes can be used by the cell to regulate its cargo traffic.

\end{abstract}



The complex internal structure of biological cells depends to a large extent on targeted transport of vesicles, organelles and other types of cargo. This active intracellular transport displays the counterintuitive property that many cargos are observed to move bidirectionally, reversing direction every few seconds \mycite{Gross04,Welte04}. This 'saltatory motion', which is faster and more persistent than Brownian motion, has been known for a long time \mycite{Rebhun67}. With the improvement of experimental techniques, bidirectional motion has been found to be widespread, including particles such as mitochondria, pigment granules, endosomes, lipid-droplets, and viruses \mycite{Welte04}.

The long-range intracellular traffic inside biological cells is powered by molecular motors which transport cargos along microtubules (MTs). Some motors such as cytoplasmic dynein walk to the minus end, while others such as kinesin~1 or 2 walk to the plus end of the MTs. Cells often have a unidirectional MT cytoskeleton: The MT minus ends are typically located near the cell center, while the plus ends point outwards towards the cell periphery. Polarized cells like epithelial cells or axons possess a unipolar parallel MT array. Because of this unidirectional nature of the MT network and the motors, both plus and minus motors must be involved in the bidirectional transport of a single cargo. Indeed both kinesin and dynein are found simultaneously on various cellular cargos \mycite{RogersGelfand97,LigonHolzbaur04,PillingSaxton06}. It is a matter of current research how the two motor species accomplish the bidirectional transport \mycite{Gross04,Welte04,KuralSelvin05,GennerichSchild06,LeviGelfand06}.

Two scenarios seem plausible \mycite{Gross04,Welte04}: (i) \textit{Tug-of-war}. Each motor species tries to move the cargo into its own direction, thereby performing a `tug-of-war' on the cargo as depicted in \FigTugOfWarCartoon. (ii) \textit{Coordination}. An additional coordination complex prevents opposing motors from being active at the same time, thereby excluding state \N\ in \FigTugOfWarCartoon. In both cases, regulatory mechanisms, which may directly target the motors or the putative coordination complex, must be present in order to allow the cell to alter its motor transport in response to internal or external stimuli.

The observed fast motion, and the complexity of bidirectional transport, as briefly reviewed in the following paragraphs, has led many authors to reject a tug-of-war scenario and search for a coordination complex. However, as shown in this article, this rejection of the tug-of-war scenario is premature since a realistic tug-of-war leads to rather complex transport behavior that is not easily understood intuitively and, thus, may be erroneously interpreted as coordinated transport.

\begin{figure}[t]\centering
\includegraphics{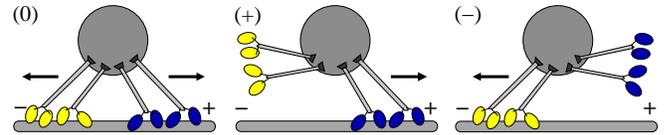}
\caption{
Cargo transport by 2 plus (blue) and 2 minus (yellow) motors: possible configurations \N, \P, and \M\ of motors bound to the MT. For configuration \N,  the motors block each other so that the cargo does not move. For configuration \P\ and \M, the cargo  exhibits fast plus and minus motion, respectively.}%
\end{figure}

Most quantitative data have been obtained experimentally in two model systems: pigment granule transport in fish and frog melanophores \mycite{NascimentoGelfand03,GrossGelfand02,LeviGelfand06} and lipid-droplet transport in \Drosophila\ embryos \mycite{WelteWieschaus98,GrossWieschaus00,GrossWieschaus02}. 
In melanophores, which are specialized pigment cells responsible for skin color, pigment granules move bidirectionally with similar velocities in both directions. They achieve net minus-end transport during an 'aggregation period' because the average distance traveled in minus direction (the minus run length) is longer than the average distance traveled in plus direction (the plus run length). During a 'dispersion period', there is almost no net transport because of an increased minus run length \mycite{GrossGelfand02}.

Lipid-droplets are storage organelles for lipids. In late \Drosophila\ embryos, they move on a unipolar MT array in the egg periphery.
Their bidirectional motion exhibits different patterns in different stages of embryonic development. In particular, from so-called phase II to III 
their net transport direction changes from plus to minus because of an increase in the minus run length \mycite{WelteWieschaus98,GrossWieschaus02}. This system is the only one for which force measurements have been performed so far. Stall forces have been found equal in plus and minus direction, independent of the net direction of droplet transport \mycite{WelteWieschaus98,GrossWieschaus02}. 

Various proteins that are necessary for the proper function or regulation of motor transport have been identified \mycite{ReileinGelfand01}. Examples are the dynein cofactor dynactin, which is necessary for bidirectional transport in melanophores \mycite{DeaconGelfand03}, or proteins like halo, klar or LDS2 in the lipid-droplet system \mycite{WelteGross05}. 

Motor transport was found to be affected both by intracellular regulation and by mutational changes in the motor structure. 
First, cellular regulation often leads to changes in only one direction. In the lipid-droplet system, net transport during embryogenesis is altered via a change in the plus run length \mycite{WelteWieschaus98}, while in the melanophore system during skin color change the minus run length is changed \mycite{GrossGelfand02}. Similarly, herpesvirus capsids achieve targeting during entry and egress by modulation of the plus run length \mycite{SmithEnquist04}. In all cases, the other direction is left unaltered. 
Second, mutation of the plus or minus motor mostly causes reduced motion in both directions by decreasing run lengths or velocities, as observed by mutating dynein on lipid-droplets \mycite{GrossWieschaus00,GrossWieschaus02} and kinesin on axonal protein carrying vesicles \mycite{KaetherDotti00}. However, in melanophores, kinesin inactivation leads to breakdown of plus motion and increased minus run lengths \mycite{GrossGelfand02}. 

Interfering with the dynein-cofactor dynactin impairs transport in both directions in melanophores \mycite{DeaconGelfand03}, but impairs minus and enhances plus transport of adenovirus particles \mycite{SuomalainenGreber99}. In the only \invitro\ experiment concerning bidirectional transport \mycite{ValeBrown92}, a motility assay of kinesin and dynein, it was observed that increasing the number of dyneins enhances minus and impairs plus end transport.

\begin{figure}[t]\centering
\includegraphics{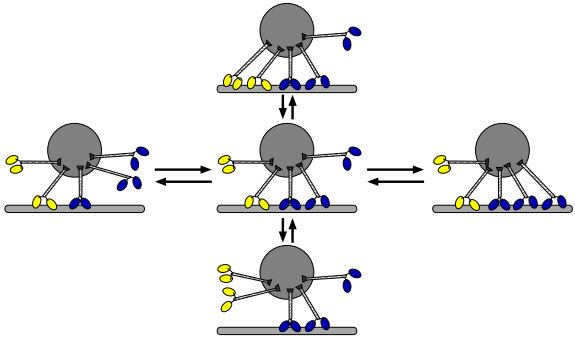}
\caption{
A cargo with $\Np=3$ plus (blue) motors and $\Nm=2$ minus (yellow) motors is pulled by a fluctuating number of motors bound to the MT. The configuration in the middle corresponds to $(\np,\nm) = (2,1)$. Only 5 out of 12 possible $(\np,\nm)$ configurations are displayed.}%
\end{figure}

As shown in this article, all of these experimental observations are consistent with the tug-of-war mechanism. In fact, we present the first explicit tug-of-war model that takes into account the experimentally known single motor properties and makes quantitative predictions for bidirectional transport.
In our model, the motors act independently and interact only mechanically via their common cargo. We find seven possible motility regimes for cargo transport. Three of these regimes are dominated by the three configurations \N, \P, and \M\ in \FigTugOfWarCartoon\ and represent no motion, fast plus motion, and fast minus motion of the cargo, respectively. The other motility states are combinations thereof; in particular, there are the two regimes \PM\ and \PMN\ where the cargo displays fast bidirectional transport without and with pauses, respectively. During fast plus or minus motion, only one motor type is pulling most of the time and the tug-of-war appears to be coordinated.

The different motility regimes are found for certain ranges of single-motor parameters such as stall force and MT affinity. Small changes in these parameters lead to drastic changes in cargo transport, \eg from fast plus motion to bidirectional motion or no motion. We propose that cells could use the sensitivity of the transport to the single-motor properties to regulate its traffic in a very efficient manner. We illustrate this general proposal by providing an explicit and quantitative tug-of-war model for the lipid-droplet system.

\section{Results}

\subsection{Model}

To study the bidirectional transport of cargos, we developed a model for a cargo to which $\Np$ plus and $\Nm$ minus motors are attached. Typically these numbers will be in the range of 1 to 10 motors as observed for many cargos  \invivo\ \mycite{AshkinSchliwa90,WelteWieschaus98,BlockerGriffiths97}. For $\Np = 0$ or $\Nm= 0$, we recover the model for cooperative transport by a single motor species as studied in \mycite{KlumppLipowsky05PNASCargoTransport}. 
We characterize each motor species by six parameters as measured in single molecule experiments (see \TabMotorParameters\ and text in the Supporting Information (SI)) as follows: it binds to a MT with the binding rate $\piO$ and unbinds with the unbinding rate $\epsO$, which increases exponentially under external force, with the force scale given by the detachment force $\Fd$. When bound to the MT, the motor walks forward with the velocity $\vF$, which decreases with external force and reaches zero at the stall force $\Fs$. Under superstall external forces, the motor walks backward slowly with backward velocity $\vB$. 

\begin{table}[t]
\caption{
Values of the single-motor parameters for kinesin~1, cytoplasmic dynein and an unknown plus motor kin? that transports \Drosophila\ lipid-droplets. The kinesin~1 values are taken from the cited references. The starred values are obtained by fitting experimental data of \Drosophila\ lipid-droplet transport and agree with the cited references.}
\begin{tabular}{@{\vrule height 10.5pt depth4pt  width0pt}l|l|l|l}\hline
parameter						& kinesin 1	&  dynein & kin?\\\hline
stall force $\Fs$ [pN]			& 6		\hfill\mycite{SvobodaBlock94,SchnitzerBlock00}
							& 1.1$^{*}$\hfill\mycite{WelteWieschaus98,MallikGross05}
							& 1.1$^{*}$\hfill\mycite{WelteWieschaus98}\\
		&					& 7		\hfill\mycite{TobaHiguchi06} &\\
detachment force $\Fd$ [pN]		& 3		\hfill\mycite{SchnitzerBlock00}
							& 0.75$^{*}$
							& 0.82$^{*}$\\
unbinding rate $\epsO$ [s$^{-1}$]	& 1		\hfill\mycite{SchnitzerBlock00,ValeYanagida96}
							& 0.27$^{*}$ \hfill\mycite{MallikGross05,KingSchroer00}
							& 0.26$^{*}$\\
binding rate $\piO$ [s$^{-1}$]	& 5		\hfill\mycite{LeducProst04}
							& 1.6$^{*}$ \hfill\mycite{KingSchroer00,ReckPetersonVale06}
							& 1.6$^{*}$\\
forward velocity $\vF$ [$\mu$m/s]	& 1		\hfill\mycite{ValeYanagida96,CarterCross05}
							& 0.65$^{*}$ \hfill\mycite{KingSchroer00,NishiuraSutoh04}
							& 0.55$^{*}$\\
back velocity $\vB$ [nm/s]	& 6	\hfill\mycite{CarterCross05}	
							& 72$^{*}$			
							& 67$^{*}$\\
\hline
\end{tabular}
\end{table}

The motors on the cargo bind to and unbind from a MT in a stochastic fashion, so that the cargo is pulled by $\np\leq\Np$ plus and $\nm\leq\Nm$ minus motors, where $\np$ and $\nm$ fluctuate with time, see \FigCargoMotorFluct. We have derived the rates for unbinding of one of the bound motors and for binding of an additional motor on the cargo from the single motor rates under the assumption that: (i) the presence of opposing motors induces a load force, and (ii) this load force is shared equally by the bound motors belonging to the same species (see SI text). We obtain a Master equation for the motor number probability $p(\np,\nm)$ that the cargo is pulled by $\np$ plus and $\nm$ minus motors. The observable cargo motion is characterized by the motor states $(\np,\nm)$ with high probability. If there is high probability for a state $(\np,0)$ or $(0,\nm)$ with only one motor species bound, corresponding to \FigTugOfWarCartoon\P\ and \M, the cargo exhibits fast plus or minus motion, respectively. If there is high probability for a state with both motor species active, \ie $\np>0$ and $\nm>0$, the cargo displays only negligible motion into the direction of the motors that 'win' the tug-of-war, because the losing motors walk backward only very slowly. This corresponds to the blockade situation depicted in \FigTugOfWarCartoon\N. 


\subsection{Motility states for the symmetric case}

We first studied the instructive symmetric case, for which the number of plus and minus motors are the same and where plus and minus motors have identical single-motor parameters except for their preferred direction of motion. Apart from being theoretically appealing, this symmetric situation can be realized \invitro\ if cargos are transported by a single motor species along antiparallel MT bundles, and can also be used \invivo\ provided plus and minus end transport exhibit sufficiently similar transport characteristics.

We solved our model for fixed motor numbers $\Np=\Nm$ and fixed single-motor parameters and determined the probability distribution $p(\np,\nm)$, see SI text. Depending on the values of these parameters, the model exhibits qualitatively different solutions, see \FigMotilityCharSym, which we will call 'motility states' in the following. These motility states exhibit distinct cargo trajectories and velocity distributions as shown in \FigMotilityCharSym\ and can formally be distinguished by the number of maxima of the motor number probability distribution $p(\np,\nm)$. This number of maxima is found to be either 1, 2, or 3. For the symmetric case, three types of maxima with the configurations of \FigTugOfWarCartoon\ occur: a maximum with only plus and no minus motors bound \P, one with only minus and no plus motors bound \M, and one with equal numbers of plus and minus motors \N. These maxima are found in the combinations \N, \PM, and \PMN, leading to three qualitatively different motility states.

\textbf{\N\ No motion} For 'weak' motors with small stall to detachment force ratio $f=\Fs/\Fd$, the probability distribution $p(\np,\nm)$ has a single maximum at a state with an equal number of bound plus and minus motors, see \FigMotilityCharSym A1, and the velocity distribution has a peak at zero velocity, see \FigMotilityCharSym A3. The corresponding cargo trajectories in \FigMotilityCharSym A2 exhibit only small fluctuations around the initial position. This motility state \N\ represents the blockade situation shown in \FigTugOfWarCartoon\N\ which one naively expects for a tug-of-war scenario.

\textbf{\PM\ Fast plus and minus motion} For strong motors with large $f$, cargo movement is completely different. The cargo switches between fast plus-directed and minus-directed motion, see \FigMotilityCharSym B2, and the probability distribution $p(\np,\nm)$ has two maxima, see \FigMotilityCharSym B1. At one maximum only plus motors are bound to the MT ($\np>0$, $\nm=0$) and at the other only minus motors ($\np=0$, $\nm>0$), corresponding to the states \P\ and \M\ in \FigTugOfWarCartoon\ which are usually associated with coordinated transport rather than with a tug-of-war scenario. 
This behavior can be understood as follows: When more plus than minus motors are bound to the MT ($\np>\nm$), every plus motor experiences the force $\FC/\np$, while every minus motor experiences the larger force $\FC/\nm$, where $\FC$ denotes the total force on the cargo. Since the unbinding rate increases strongly with increasing load force, minus motors are more likely to unbind from the MT than plus motors, so that the predominance of the plus motors is further enhanced. After the unbinding of a minus motor the remaining minus motors experience an even larger force and are even more likely to unbind. As a consequence, the cargo experiences a cascade of minus motor unbinding events until no minus motor remains bound.
A prerequisite for this unbinding cascade is that the motors can exert a sufficiently large force to pull off opposing motors from the MT, \ie the stall force $\Fs$ has to be comparable or larger than the detachment force $\Fd$.
For small force ratios $f=\Fs/\Fd$, the pulling force has only a small effect on motor unbinding, so that no instability occurs and the cargo exhibits the blocked motility state \N. For large motor force ratio, the transient predominance of one motor type is thus amplified by a dynamic instability and most of the time only one motor type is bound, as indicated in  \FigTugOfWarCartoon\P\ and \M. 
The emergence of cooperative behavior arising from the nonlinear force dependence of the unbinding rate has also been proposed as an explanation for collective effects in muscles \mycite{Duke00} and mitotic spindle oscillations \mycite{GrillJulicher05}.
For the tug-of-war of 4 against 4 motors with kinesin-like parameters as in \FigMotilityCharSym B, about $90\%$ of the time only one motor type is bound. During a plus or minus run, the effective velocity is however slightly reduced compared to the single-motor velocity by the sporadic binding and subsequent fast unbinding of an opposing minus motor. The velocity distribution in \FigMotilityCharSym B3 has two peaks close to the single-motor velocities $\pm 1\,\mu$m/s.
The direction of motion of the cargo is reversed when, due to a fluctuation, the defeated motors become predominant.

\begin{figure}\centering
\includegraphics{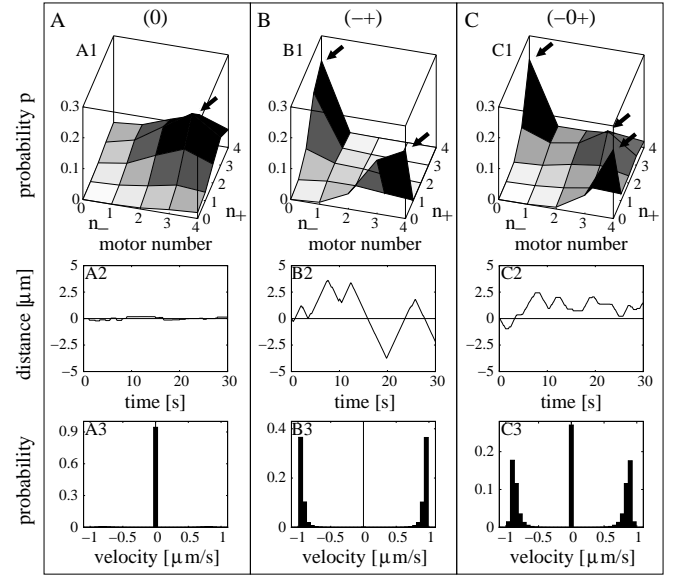}
\caption{
Motility states for the symmetric tug-of-war of $\Np=\Nm=4$ plus and minus motors. The three columns (A), (B) and (C) correspond to the three motility states \N, \PM\ and \PMN: 
(A) The no-motion motility state \N\ is characterized by (A1) motor number probabilities $p$ with a single maximum at an equal number of active plus and minus motors, (A2) trajectories with almost no motion, and (A3) velocity distributions with a single maximum at zero velocity.
(B) The motility state \PM\ of fast bidirectional motion is characterized by (B1) probabilities $p$ with two maxima with only plus or only minus motors active, (B2) trajectories which exhibit switching between fast plus and minus motion, and (B3) bimodal velocity distributions with two peaks close to the single-motor velocities of $\pm$ 1 $\mu$m/s.
(C) The motility state \PMN\ is characterized by (C1) probabilities with three maxima corresponding to fast plus and minus and no motion, (C2) trajectories which exhibit fast plus and minus motion and pauses, and (C3) velocity distributions with three peaks. 
Both plus and minus motors in (B) have the kinesin~1 parameters of \TabMotorParameters. The different motility behaviour in (A) and (C) is obtained by changing the single motor parameters in \TabMotorParameters to (A) $\Fs$ = 2 pN, and (C) $\Fs$ = 4.75 pN, and $\epsO$ = 0.4 s$^{-1}$.%
}%
\end{figure}

\textbf{\PMN\ Fast plus and minus motion with interspersed pauses} Finally, in some intermediate parameter ranges, the probability distribution $p(\np,\nm)$ exhibits three maxima as shown in \FigMotilityCharSym C1, a symmetric one corresponding to no motion as for motility state \N\ and two nonsymmetric ones corresponding to steady plus and minus motion as for state \PM. The velocity distribution has three corresponding peaks, see \FigMotilityCharSym C3, and cargo trajectories therefore exhibit bidirectional motion interspersed with pauses, see \FigMotilityCharSym C2.


\subsection{Motility states for the asymmetric case}

Bidirectional cargo transport \invivo\ is typically dependent on two different motor species for plus and minus motion. This plus-minus asymmetry can lead to net transport of the cargo in one direction. For example, in the motility state \PMN, the plus motion maximum \P\ of the motor number probability can be larger than the minus motion maximum \M, see \FigMotilityCharAsym A1, which leads to longer plus runs compared to minus runs and to net plus motion of the cargo as illustrated by the trajectory in \FigMotilityCharAsym A2. The velocity distribution in \FigMotilityCharAsym A3 has the three peaks characteristic for the \PMN\ regime but the peak at high plus velocity is larger than the one at high minus velocity. As cargo motion is no longer symmetric with respect to plus and minus motion, seven motility states are now possible, corresponding to the different combinations \P, \M, \N, \PM, \PN, \NM and \PMN\ of the maxima \P, \M, and \N. The new motility states  \PN\ and \P\ are shown in \FigMotilityCharAsym B, C. The two other new states \NM\ and \M\ are analogous with plus and minus motion interchanged. 

In the motility state \PN, the motor number probability has a maximum at the plus motion state \P\ with only plus motors active and a maximum at the no-motion state \N\ with both types of motors active, see \FigMotilityCharAsym B1. The corresponding velocity distribution in \FigMotilityCharAsym B3 has two peaks, one close to zero velocity and one at large plus motor velocity, and the cargo switches between fast plus motion and pauses, see \FigMotilityCharAsym B2. In the \P\ motility state in \FigMotilityCharAsym C, the motor number probability and velocity distribution exhibit a maximum corresponding to fast plus motion\footnote{The small peak near zero velocity corresponds to the no-motion states near the maximum for which both $\np$ and $\nm$ are non-zero.}. 


\subsection{\Invivo\ tug-of-war}

To check whether our model can account quantitatively for experimental observations, we applied our model to the bidirectional movements of lipid-droplets in \Drosophila\ embryos. We chose this particular series of sophisticated experiments \mycite{WelteWieschaus98,GrossWieschaus00,GrossWieschaus02} because it is unique in providing an estimate for the number of motors on the cargo, a high number of quantitative measurements of transport characteristics including cargo force measurements, as well as observations in two different developmental phases (labeled wild type phase II and III, \WtII\ and \WtIII) and in three different dynein mutation backgrounds. The droplets are transported by an unknown plus motor, presumably an unconventional kinesin, and cytoplasmic dynein \mycite{GrossWieschaus00}.

\begin{figure}\centering
\includegraphics{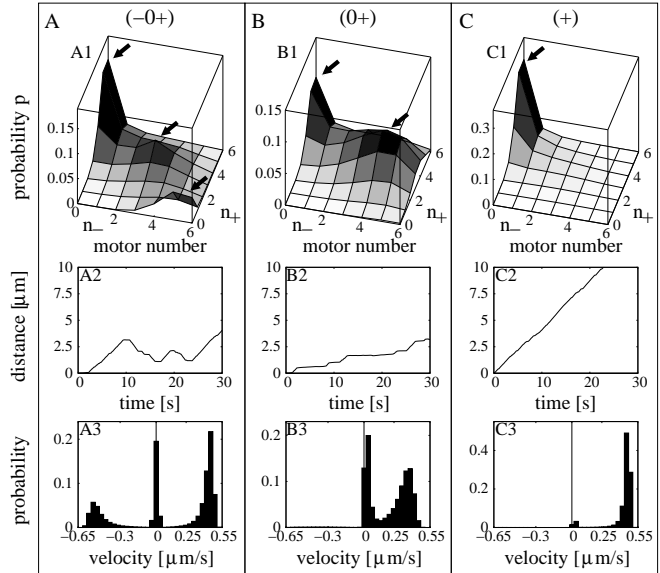}
\caption{
Motility states for the asymmetric tug-of-war of $\Np=6$ plus against $\Nm=6$ different minus motors. 
The cargo is in one of seven motility states. The motility states \N, \PM, and \PMN\ are as for the symmetric case shown in \FigMotilityCharSym, except that the plus-minus symmetry is lost as illustrated in column (A) for the \PMN\ motility state. 
(A) The \PMN\ motility state is characterized by (A1) three maxima in the motor number probability $p$ at a plus, a minus and a no-motion state, (A2) trajectories with rapid plus and minus motion interspersed with pauses, and (A3) three peaks in the velocity distribution. Plus motion has a higher probability so that net motion is plus-end directed.
(B) The \PN\ motility state is characterized by (B1) probabilities $p$ with one maximum with only plus motors and one with plus and minus motors active, (B2) trajectories with fast plus motion and pauses, and (B3) velocity distributions with two peaks near the single plus motor velocity $\vFP$ = 0.55 $\mu$m/s and near zero.
(C) The \P\ motility state is characterized by (C1) probabilities $p$ with a maximum with only plus motors active, (C2) trajectories with fast plus motion, and (C3) velocity distributions with a peak close to the single plus motor velocity.
The motility states \NM\ and \M\ are similar to the states \PN\ and \P\ with plus and minus interchanged.
(A) represents lipid-droplet transport: The plus and minus motors have the \Drosophila\ plus motor (kin?) and dynein parameters of \TabMotorParameters. The same parameters are used in (B) and (C) except for (B,C) $\FsM$ = 0.45 pN  and (B) $\epsOM$ = 0.24 s$^{-1}$ and (C) $\epsOM$ = 0.54 s$^{-1}$.}%
\end{figure}

We first considered the \WtII\ data. Cargo stall force measurements led to the conclusion that the droplets are on average pulled by 5 plus and 5 minus motors, and that both types of motors have a single-motor stall force of $1.1\,$pN \mycite{WelteWieschaus98}. Since the number of active motors fluctuates stochastically, this should be the average number of pulling motors. We therefore fixed the total number of plus and minus motors to $\Np=\Nm=6$. 

We then performed simulations and varied the undetermined single motor parameters in order to fit the experimentally measured transport characteristics, namely plus and minus run lengths, plus and minus stall forces, pause times after plus and minus travel, and plus and minus velocities of short and long runs, with an accuracy of ca. 10$\%$ (for the detailed procedure and results of this and the following fits see SI text and Table 2 and 3). The resulting parameters for dynein and the unknown plus motor (kin?) are listed in \TabMotorParameters. The dynein parameters are in agreement with \invitro\ measurements of dynein properties when available. All other parameter lie in a reasonable range. The dynein backward velocity is an order of magnitude larger than for kinesin~1, in agreement with experiments \mycite{MallikGross05,WangSheetz95}.

\FigMotilityCharAsym A shows a sample trajectory, the motor number probability and the velocity distribution for the droplet tug-of-war in \WtII. The cargo switches between fast plus and minus motion and pauses but exhibits net plus motion because the probability for $\P$ states is higher than for $\M$ states. The cargo stall forces in plus and minus direction are equal, see SI \TabFitResults. This shows that the cargo direction is not only determined by the motor forces but also by other motor properties, see \TabMotorParameters. In this case, the higher plus motor detachment force makes it difficult to rip off the plus motors and thus favors plus motion.

\begin{figure*}\centering
\includegraphics{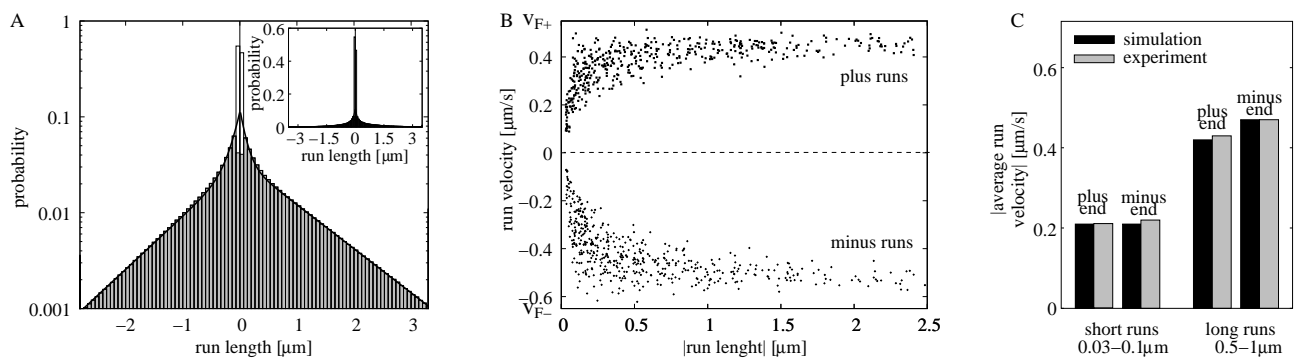}
\caption{
\Drosophila\ lipid-droplet transport in wild type phase II: tug-of-war of 6 \Drosophila\ plus motors and 6 dyneins with parameters as in \TabMotorParameters. %
(A) Distribution of run length, the distance traveled in one direction before a pause or a directional switch occurs. Minus (plus) run lengths are negative (positive). The gray bars are the run lengths observable with the experimental cutoffs of a minimum length of 0.16 s and 30 nm, while the white bars are obtained without the cutoffs and thus beyond experimental resolution. The lines are double exponential fits to the simulation data with decay lengths of ca. 0.1 $\mu$m and 1 $\mu$m in both directions, of the same order of magnitude as in the experiments \mycite{GrossWieschaus02}.
(B) Scatter plot of the run velocity (positive for plus and negative for minus runs) versus the run length of 500 runs in each direction shows a positive correlation: longer runs have higher velocities. The Spearman rank correlation coefficient both for plus and minus motion is larger than 0.7 with a significance level below 10$^{-10}$. Long runs have almost the maximal velocity which is the single motor velocity, $\vFP$ in plus and $-\vFM$ in minus direction. There are no data points for small run lengths and velocities because runs have been defined as periods with a velocity larger than 50 nm/s for at least 30 nm.
(C) The correlation of run length and velocity can also be seen by considering short (0.03-0.1 $\mu$m) and long (0.5-1 $\mu$m) plus end \resp minus end runs. Short runs have lower averages than long runs, which reproduces the experimental averages of \mycite{GrossWieschaus02}.%
}%
\end{figure*}

A nontrivial consistency check of our model is provided by three additional features that we obtained from this model in close agreement with experimental observations even though these features were not used in order to determine the model parameters in \TabMotorParameters. 
First, the distribution of plus and minus run lengths can be fitted by a sum of two exponentials\footnote{This has also been found in the melanophore system \mycite{GrossGelfand02}.} with length scales of the same order of magnitude as obtained experimentally \mycite{GrossWieschaus02}, see \FigMotilityWtII A.
Second, the pause time distributions of pauses after plus and after minus runs are very similar and can be fitted by a single exponential function with a time scale of the same order of magnitude as in the experiments \mycite{GrossWieschaus00}, see SI \FigWtIIPauseTimes.
Third, there is a correlation between run length and run velocity: long runs have larger average velocities, see \FigMotilityWtII B. In the experiments \mycite{GrossWieschaus00,GrossWieschaus02}, this has been quantified by dividing the runs into short runs ($0.03-0.1\,\mu$m) and long runs ($0.5-1\,\mu$m). Short runs have approximately half the velocity of long runs, see  \FigMotilityWtII C. In our model, this property reflects the correlation of the average number of active winning motors with the run length, see SI \FigCorrVelRunLength, and can be understood as follows.

During a certain run, \eg in the plus direction, minus motors bind from time to time to the MT. This slows down the motion and causes a 'pause'. However, the active plus motors generate a large force on this single minus motor, which is then ripped off fast from the MT. As a consequence, the 'pauses' are too short to be detected experimentally and are only noticeable via the decreased average cargo velocity in the plus direction. If a cargo is pulled by many plus motors, this has two effects: (i) The effective cargo velocity is increased because opposing minus motors do not create large forces on each of the many plus motors and because the minus motor drops off very fast. (ii) The plus run length is larger because it is less probable that the minus motors take over. Both effects together lead to a correlation of run length and velocity.


\subsection{Mutation and regulation}

Three different dynein mutations in \Drosophila\ embryos of developmental phase II have been studied in \mycite{GrossWieschaus00,GrossWieschaus02},  and all three lead to impairment of both plus and minus motion with reduced run lengths and stall forces. At first sight, this simultaneous impairment of both transport directions in response to mutations that affect only one of the two motor species seems to be inconsistent with a tug-of-war. However, using our tug-of-war model, we were again able to describe the observed behavior with an accuracy of about  10\%. In order to do so, we only varied the minus motor paramaters and kept the plus motor parameters fixed to their \WtII\ values, see SI \TabFitResults.

In our model, the dynein mutations simultaneously modify several parameters of this motor, among which are its unbinding rate, its binding rate, and its detachment force. If only one of these parameters were modified, the resulting motor behavior would be easy to understand. First, if only the unbinding rate is increased, the minus motors unbind from the filament faster and thus produce less force on the plus motors, which leads to longer plus and shorter minus runs. Second, increasing only the minus motor binding rate  has the opposite effect because dyneins are more likely to rebind to the filament. Third, if  only the minus motor detachment force is enhanced, the ability of the minus motors to resist the plus motors is also enhanced, which increases minus and decreases plus run lengths. Therefore, if only a single parameter of the minus motor is modified,  motion in one direction is enhanced whereas motion in the opposite direction is impaired. On the other hand, the overall effect of changes in several motor parameters is difficult to anticipate intuitively
and can lead to impairment of both directions as shown in our model. 

Furthermore, two different embryonic phases \WtII\ and \WtIII\ allow to assess the effect of cellular regulation. In \WtII\ net droplet transport is plus-end directed, while it is minus-end directed in \WtIII\ due to a reduction in plus run lengths. Apart from the stall forces, all other transport characteristics remain unchanged. We propose that the cellular regulation that causes this change targets the motor properties. We therefore fitted the \WtIII\ data by varying the single-motor parameters as for the \WtII\ data. The fit shows that a tug-of-war can lead to impairment of motion in one direction while leaving the other direction unaffected, see SI \TabFitResults. The obtained single motor parameters for \WtII\ and \WtIII\ are rather similar. This sensitivity of motion to the single-motor parameters allows the cell to regulate its transport in an efficient way.

In the only \invitro\ experiment concerning bidirectional transport \mycite{ValeBrown92}, a motility assay of kinesin and dynein, it was observed that increasing the number of transporting dyneins enhances minus and impairs plus end transport. This is reproduced in our model, compare SI \TabFitResults.


\section{Discussion}

We have investigated a model for bidirectional cargo transport based on a tug-of-war between plus and minus motors governed by mechanical interactions only. Our model exhibits many features that are usually attributed to a coordination mechanism. In particular, even for equally strong motors, a tug-of-war does not necessarily lead to the expected blockade situation with almost no cargo motion as in \FigTugOfWarCartoon\N, but can also lead to switching between fast plus end and fast minus end motion as in \FigTugOfWarCartoon\P\ and \M. This surprising behavior is caused by a dynamic instability arising from the strongly nonlinear force-dependence of the single-motor unbinding rate. This instability leads to a high probability of having only one motor type active at a given time.

In our tug-of-war model, the motility behaviour of the cargo is very sensitive to the single motor properties. Changing the motor stall force or MT affinity, for example, can lead to qualitatively different motility behaviour such as fast plus motion, no motion, or bidirectional transport. When we modified the single-motor properties to mimic the effect of either mutations or of regulatory processes, we found that motion in plus and minus direction can be affected in various ways. We found cases for which (i) motion was affected only in one direction, (ii) motion was impaired in one direction and enhanced in the other, and (iii) motion was enhanced or impaired in both directions. This variability agrees with experimental observations where different systems also exhibit widely varying reactions to regulation or mutation \mycite{GrossGelfand02,WelteWieschaus98,GrossWieschaus00,GrossWieschaus02,DeaconGelfand03,SmithEnquist04,KaetherDotti00,SuomalainenGreber99} as described in the introduction. 

Our tug-of-war model is thus in qualitative agreement with experimental data for bidirectional transport \invivo. Furthermore, we have been able to quantitatively describe the experimental data for the \Drosophila\ lipid-droplet system. The latter system exhibits different transport regimes depending (i) on the different phases of the embryonic development, which are distinguished by distinct sets of regulatory proteins, and (ii) on the molecular structure of the motor proteins, which have been changed by mutations. In our theory, these different transport regimes arise from variations in single motor parameters, but the basic transport mechanism underlying all of these regimes is provided by a tug-of-war between the two motor species.

Our results show that the two scenarios for bidirectional transport displayed in \FigTugOfWarCartoon, namely the tug-of-war and coordinated motor activity, are not mutually exclusive, but rather that the tug-of-war provides a mechanism for coordinated movement.\\


\begin{acknowledgments}
We thank C{\'e}cile Leduc for pointing out a problem in a previous version of our model. SK was supported by Deutsche Forschungsgemeinschaft (Grants KL818/1-1 and 1-2) and by the National Science Foundation through the Physics Frontiers Center-sponsored Center for Theoretical Biological Physics (Grants PHY-0216576 and PHY-0225630).
\end{acknowledgments}


\newpage

\section{\huge Supporting information}
\vspace{0.7\baselineskip}
\textbf{for M.J.I. M{\"u}ller, Stefan Klumpp, Reinhard Lipowsky: Tug-of-war as a cooperative mechanism for bidirectional cargo transport by molecular motors}
\vspace*{\baselineskip}

\setcounter{footnote}{0}

\section{Model for a single motor}

A single motor can walk along a MT with velocity $v$, unbind from the MT with unbinding rate $\eps$ and bind to the MT with binding rate $\pi$. Our choice of the rates is based on the load-dependent transport parameters of single motors as measured in single-molecule experiments.
When bound to the MT, the motor moves forward with a load-dependent velocity which decreases monotonically from the zero-load forward velocity $\vF$ to zero at the stall force $\Fs$ \mycite{SvobodaBlock94A,KojimaYanagida97A,VisscherBlock99A,TobaHiguchi06A}. For superstall loads the motor walks slowly backwards, as has been shown for kinesin \mycite{KojimaYanagida97A,CarterCross05A}. In our model, we used a piecewise linear force-velocity relation with 
\begin{eqnarray}\label{eq:velocitySingleMotor}
v(F) = \left\{\begin{array}{lrl}
	\vF\left(1-F/\Fs\right) &\mbox{for} & F\leq\Fs\\
	\vB\left(1-F/\Fs\right)&\mbox{for} & F\geq\Fs
\end{array}\right.
\end{eqnarray}
Here, $\vB$ is the absolute value of the motor backward velocity. For forces smaller than the stall force, such a linear relation provides a good approximation for the experimentally determined force-velocity curves both for kinesin \mycite{SvobodaBlock94A,KojimaYanagida97A,VisscherBlock99A} and dynein \mycite{TobaHiguchi06A}. For superstall forces, the shape of the force-velocity curve is not known precisely. In this range our linear relation can be considered as a Taylor expansion to first order in $F-\Fs$. The detailed form of the force-velocity curve is however not crucial for our results, as long as it decreases monotonously and exhibits a small backward velocity. The unbinding rate of the motor from the MT increases exponentially with the applied force $F$:
\begin{eqnarray}\label{eq:unbindingRateSingleMotor}
	\eps(F)=\epsO\,\exp\left(|F|/\Fd\right)
\end{eqnarray}
as measured for kinesin \mycite{SvobodaBlock94A} and as follows from Kramers or Bell theory. The force scale is set by the detachment force $\Fd$.
The binding rate to the MT is difficult to assess experimentally. It is expected to depend only weakly on the load because an unbound motor relaxes and then binds from its relaxed state (see the discussion in \mycite{KlumppLipowsky05PNASCargoTransportA}). We therefore take the binding rate equal to the zero-load binding rate $\piO$, independent of load:
\begin{eqnarray}\label{eq:bindingRateSingleMotor}
	\pi(F)=\piO
\end{eqnarray}

The single motor rates of kinesin~1, cytoplasmic dynein and an unknown plus motor that transports lipid-droplets in \Drosophila\ embryos are shown in \TabMotorParameters. For kinesin~1 all parameters have been measured in single-molecule experiments, see the references in the table. 
For dynein, only part of the parameters have been measured, and for the stall force conflicting results have been reported by different labs, see the references given in \TabMotorParameters. In addition, dynein is very sensitive to regulatory and accessory proteins \mycite{MallikGross04RevA}.
The unknown dynein parameters and the parameters of the unknown \Drosophila\ plus motor are obtained by fitting experimental data from \Drosophila\ droplet transport \mycite{WelteWieschaus98A,GrossWieschaus00A,GrossWieschaus02A} as described in the section 'Fit to the lipid-droplet data' below.

\section{Effective rates for the cargo}

We consider a cargo that is transported by constant numbers of $\Np$ plus and $\Nm$ minus motors. At every time $t$, the state of the cargo with $\Np$ plus and $\Nm$ minus motors firmly attached to it is fully characterized by the numbers $\np$ and $\nm$ of plus and minus motors that are bound to the MT and thus actively pull on the cargo at that time. The state of the cargo changes when a plus or a minus motor binds or unbinds to/from the MT, see \FigCargoMotorFluct. These changes are described by a Master equation for the probability distribution $p(\np,\nm,t)$ to have $\np$ bound plus and $\nm$ bound minus motors at time $t$. The rates of this Master equation describe the transitions corresponding to the arrows in \FigCargoMotorFluct\ and are determined from the single-motor rates using the assumptions that the motors act independently and feel each other only due to two effects: (i) opposing motors act as load, and (ii) identical motors share this load. If each plus motor feels the load $\Fp$ (and generates the force $-\Fp$) and each minus motor feels the load $-\Fm$ (and generates the force $\Fm$), this means that the force balance on a cargo pulled by $\np$ plus and $\nm$ minus motors is 
\begin{eqnarray}\label{eq:forceBalance}
\np\Fp = -\nm\Fm \equiv \FC.
\end{eqnarray}
Here, the sign of the force is chosen positive if it is a load on the plus motors, \ie if it points into the minus direction. If only one motor type is bound, \ie if $\np=0$ or $\nm=0$, then $\Fp=\Fm=\FC=0$. A single bound plus motor thus feels the force $\Fp=\FC/\np$. Using \EqsAnd{eq:unbindingRateSingleMotor}{eq:bindingRateSingleMotor}, this implies that the effective rate for the unbinding of one plus motor is 
\begin{eqnarray}\label{eq:cargoEffectiveUnbinding}
\np\epsOP\,\exp\left[\FC/(\np\FdP)\right],
\end{eqnarray}
and the effective rate for the binding of one plus motor is 
\begin{eqnarray}\label{eq:cargoEffectiveBinding}
(\Np-\np)\,\piOP.
\end{eqnarray}
Here and in the following, the index '$+$' labels plus motor properties. Analogous expressions hold for the minus motors with the parameters indexed by '$-$'. 

The cargo force $\FC$ is determined by the condition that plus motors, which experience the force  $\FC/\np$, and minus motors, which experience the force $-\FC/\nm$, move with the same velocity, which is the cargo velocity $\vC$:
\begin{eqnarray}\label{eq:velocityBalance}
	\vC(\np,\nm) = \vP\left(\FC/\np\right) = -\vM\left(-\FC/\nm\right)
\end{eqnarray}
Here, the sign of the velocity is taken positive in the plus direction and negative in the minus direction. 
In order to have a unique solution $\FC$ to this equation, both motors must have nonzero backward velocities; otherwise the single-motor force velocity relations $\vP(F)$ and $\vM(F)$ do not have well-defined inverses.
In the case of 'stronger plus motors' $\np\FsP>\nm\FsM$, \EqsAnd{eq:velocitySingleMotor}{eq:velocityBalance} lead to the cargo force and velocity
\begin{eqnarray}
    \FC(\np,\nm) &=& \lambda\,\np\FsP + (1-\lambda)\,\nm\FsM \label{eq:cargoForce}\\
    \vC(\np,\nm) &=& \frac{\np\FsP-\nm\FsM}{\np\FsP/\vFP+\nm\FsM/\vBM}\label{eq:cargoVelocity}
\end{eqnarray}
with $\lambda=1/\left(1+\left(\np\FsP\vBM\right)/\left(\nm\FsM\vFP\right)\right)$. The cargo moves to the plus direction, $\vC>0$. In the opposite case of 'stronger minus motors' with $\np\FsP<\nm\FsM$, in \EqsAnd{eq:cargoForce}{eq:cargoVelocity} the plus motor forward velocity $\vFP$ has to be replaced by its backward velocity $\vBM$, and the minus motor backward velocity $\vBM$ by its forward velocity $\vFM$. The cargo moves into the minus direction, $\vC<0$.
Typically the backward velocity is rather small, so that a cargo with $\np, \nm>0$ pulled by both types of motors usually moves very slowly, as indicated by the 'blockade' situation in \FigTugOfWarCartoon\N. If however only one motor type is bound, \eg if $\nm=0$, the cargo moves fast with the single plus motor velocity $\vC=\vFP$, corresponding to \FigTugOfWarCartoon\P.

\Eqss{eq:cargoEffectiveUnbinding}{eq:cargoEffectiveBinding}{eq:cargoForce}, and the corresponding equations for minus motors, fully determine the rates that enter the Master equation for the motor number probability $p(\np,\nm,t)$ on the state space $0\leq\np\leq\Np$, $0\leq\nm\leq\Nm$.
In each state $(\np,\nm)$, the cargo moves with velocity $\vC(\np,\nm)$ as given by \Eq{eq:cargoVelocity}. 

Experiments usually observe only cargos that have been bound to the MT for an unknown time period and monitor them over a timescale of minutes which is large compared to the times scales of motor (un-)binding, which are of the order of seconds. We are therefore interested in the long-time behaviour of the cargo which corresponds to the time-independent steady state solution $p(\np,\nm)$ of the Master equation.

\section{External forces}

If an external force $\Fext$ is present, the force balance \Eq{eq:forceBalance} becomes
\begin{eqnarray}\label{eq:forceBalanceFext}
\np\Fp = -\nm\Fm + \Fext.
\end{eqnarray}
Here again, forces are taken to be positive if they point into the minus direction. Carrying through the same calculation as for the case without external force leads to the velocity of a cargo transported into the plus direction by $\np$ active plus and $\nm$ active minus motors under an opposing external force $\Fext$: 
\begin{eqnarray}\label{eq:cargoVelocityFext}
    \vC(\np,\nm,\Fext) = \frac{\np\FsP-\nm\FsM-\Fext}{\np\FsP/\vFP+\nm\FsM/\vBM}
\end{eqnarray}
In the case of minus motion under an opposing force $\Fext$ (which is then negative), the plus motor forward velocity $\vFP$ has to be replaced by its backward velocity $\vBM$, and the minus motor backward velocity $\vBM$ by its forward velocity $\vFM$.\footnote{The case of assisting force is not treated here. It needs a definition of the single motor force-velocity relation \Eq{eq:velocitySingleMotor} also for assisting loads $F<0$.} Thus the stall force of a cargo pulled by $\np$ plus and $\nm$ minus motors is given by
\begin{eqnarray}\label{eq:cargoStallForce}
    \FsC(\np,\nm) = \np\FsP-\nm\FsM,
\end{eqnarray}
as intuitively expected.


\section{Numerical calculations}

For a given set of single-motor parameters, we numerically calculate the steady state solution $p(\np,\nm)$ as the nullspace of the transition matrix of the Master equation \mycite{vanKampen92A}. Such motor number probabilities are shown in \FigMotilityCharSym A2-C2 and \FigMotilityCharAsym A2-C2.
We then determine the locations $(\nptilde,\nmtilde)$ of maxima of $p(\np,\nm)$ which define the 'motility state' of the cargo. $p(\nptilde,\nmtilde)$ is a maximum if it is larger than its direct and diagonal neighbours on the state space $(\np,\nm)$. A maximum at a state $(\nptilde,0)$ with $\nptilde>0$ corresponds to a plus motion state labeled by '$+$', a maximum at $(0,\nmtilde)$ with $\nmtilde>0$ to a minus motion state labeled by '$-$', and a maximum at a state $(\nptilde,\nmtilde)$ with both $\nptilde$ and $\nmtilde$ larger than zero to a no-motion state labeled by '0'. For a given set of single-motor parameters, there is at most one of each type of maximum. Thus there are seven possible combinations of maxima, which give the seven motility states \P, \M, \N, \PN, \NM, \PM, and \PMN. In the motility state \PN, for example, there are two maxima, one at a plus-motion state $(\nptilde,0)$, and one at a no-motion state $(\nptilde,\nmtilde)$. If the probability maximum is at $(0,0)$, the cargo is in the 'unbound' state.

In order to obtain dynamical quantities such as trajectories, run lengths and run velocities, we generate individual cargo trajectories using the Gillespie algorithm \mycite{Gillespie76A} for the binding/unbinding dynamics and let the cargo move with the velocity $\vC$ in the intervals between binding/unbinding events. At the start of the simulation, a cargo is bound to the filament by a random number of $\np$ active plus and $\nm$ active minus motors. In order to suppress transient behavior due to initialization bias, measurement of run lengths and velocities is started only after a time lapse of $10^4\,$s. The obtained velocity and run length distributions as shown in \FigMotilityCharSym A3-C3, \FigMotilityCharAsym A3-C3 and \FigMotilityWtII A, and all average values, are obtained from 20-50 trajectories, each of which lasts $10^6\,$s.

For comparison with experiments, the analysis of the simulated trajectories is performed in close analogy with the analysis of the experimental trajectories of \mycite{WelteWieschaus98A,GrossWieschaus00A,GrossWieschaus02A}. The cargo displacement (as shown in \FigMotilityCharSym A2-C2 and \FigMotilityCharAsym A2-C2) is recorded at video frequency of 30/s. The obtained trajectories are segmented into runs and pauses by using the definitions from \mycite{GrossWieschaus00A,GrossWieschaus02A}: A cargo is considered to be moving into the plus (minus) direction if its velocity $\vC$ is larger than 50 nm/s (smaller than $-$50 nm/s) and pausing else. A run has to be at least 30 nm and 0.16 s long, and a pause must be longer than 0.23 s and cover a distance smaller than 30 nm. The run velocity is defined as the ratio of run length and run time. 


\section{Motility diagrams}

In the symmetric tug-of-war, for which the number of plus and minus motors are the same and for which plus and minus motors have identical single-motor parameters except for their preferred direction of motion, the cargo motion depends on four dimensionless parameters: the number $N=\Np=\Nm$ of plus and minus motors, the stall force to detachment force ratio $f=\Fs/\Fd$, the MT desorption constant $K=\epsO/\piO$, and the backward-forward velocity ratio $\nu=\vB/\vF$. Depending on the values of these parameters, the cargo is in one of three possible motility states: \N, \PM, and \PMN. These states are characterized by qualitatively different motility behaviors as illustrated in \FigMotilityCharSym\ for three sets of parameters.


Regulation of cargo motion requires change of the motor parameters. To show the effect of parameter changes explicitly, we calculate the 'motility diagram' shown in  \FigMotDiagSym\ for the symmetric tug-of-war with $N=\Np=\Nm=4$ symmetric plus and minus motors. This is done as follows: The single-motor parameters are taken to be equal to the kinesin~1 values as given in \TabMotorParameters\ except for the plus and minus motor unbinding rates $\epsO=\epsOP=\epsOM$ and the stall forces $\Fs=\FsP=\FsM$. The parameter space $(\epsO,\Fs)$ is then explored systematically\footnote{All other single-motor parameters are kept constant. However, as far as backward motion is concerned, not the backward velocity $\vB$ is kept constant but rather the backward slope $\vB/\Fs$ of the force-velocity curve, with $\Fs$ equal to the kinesin~1 value.}, and for each point the maxima of the motor number probability $p(\np,\nm)$ is calculated and the cargo motility state is determined, as described in the section 'Numerical calculations' above. When the maxima between two scanned points change, we zoom in between these points in order to determine the transition point more accurately. The lines shown in \FigMotDiagSym\ consist of these points at which the locations of the maxima change.

For large desorption constants $K$, the motors have a low affinity to the MT; therefore the number of bound motors in \FigMotDiagSym\ is low for low $K$. For very high desorption constants $K$ larger than the number of motors $N=4$, the cargo is 'unbound', \ie the maximum of the motor number probability is at $(\nptilde,\nmtilde)=(0,0)$. For small force ratios $f$, the probability distribution $p(\np,\nm)$ has a single maximum at a state $(\ntilde,\ntilde)$ with an equal number $\ntilde=\nptilde=\nmtilde$ of active plus and minus motors and is in the no-motion motility state \N\ (green). For large force ratios $f$, the motors can generate forces large enough to rip off opposing motors since the stall force is large compared to the detachment force. This leads to the unbinding cascade described in the main text, and the motor number probability has maxima at states with only one active motor type, \ie at $(\nptilde,0)$ with only active plus motors and at $(0,\nmtilde)$ with only active minus motors. In the latter situation, the cargo is in the \PM\ motility state (yellow). For intermediate values of $f$, both types of maxima coexist, and the cargo is in the \PMN\ motility state (red). 

The lines in \FigMotDiagSym\ that separate regions of different color mark the parameter sets where maxima of the motor number appear or disappear and correspond to bifurcation lines. For example, when passing from  the cross labeled A in the \N\ regime with a maximum at $(\nptilde,\nmtilde)=(3,3)$, corresponding to no motion, into the \PMN\ regime by increasing the force ratio $f$, two new maxima at $(\nptilde,\nmtilde)=(4,0)$ and $(\nptilde,\nmtilde)=(0,4)$ appear. This means that the cargo now exhibits fast plus and minus motion. However, this fast motion is rarely observed since the two new maxima have less probability than the pause-maximum at (3,3). With further increase of $f$, the maxima at (4,0) and (0,4) grow, while the maximum at (3,3) shifts to (2,2) and diminishes. Cargo motion becomes more and more dominated by fast plus and minus motion, and pauses become less frequent until they finally disappear together with the maximum at (2,2) when passing the line to the \PM\ regime. At the cross labeled C in \FigMotDiagSym, the cargo switches between fast plus and minus motion. A direct transition from the \N\ to the \PM\ regime (without passing the \PMN\ regime) occurs only for maxima with low motor numbers, when the no-motion maximum  and the two fast motion maxima are neighbours (either direct or diagonal) on the discrete state space $(\np,\nm)$. This happens, for example, when passing from a region with a maximum at $(\ntilde,\ntilde)=(1,1)$ to a region with maxima at $(\ntilde,0)=(1,0)$ and $(0,\ntilde)=(0,1)$. The bifurcation line then marks the parameter set where all maxima have the same probability.

We investigated the average times between direction reversals in the \PM\ motility state and found it to grow exponentially with the motor number. This indicates that the transitions between the different motility states become nonequilibrium phase transitions in the limit of large motor numbers.

\FigMotDiagAsym\ shows the motility diagram for the tug-of-war of $\Np=6$ plus against $\Nm=6$ minus motors with parameters closely related to the \Drosophila\ lipid-droplet transport in wild type phase II. The diagram does not show maxima locations but only the bifurcation lines between the different motility states. It is generated in a similar way as \FigMotDiagSym: The single-motor parameters correspond to the values of the \Drosophila\ plus motor kin? and dynein as given in \TabMotorParameters\ except for the minus motor unbinding rate $\epsOM$ and stall force $\FsM$. The parameter space $(\epsOM,\FsM)$ is then explored systematically in the same way as for the symmetric case. \FigMotDiagAsym\ exhibits seven motility states. The 'new' motility states \P, \M, \PN\ and \NM\ are asymmetric with respect to plus and minus motors and, thus, were not present for the symmetric tug-of-war. \FigMotDiagAsym\ shows that minus motors with high affinity to the MT, \ie with low desorption constant $\KM$, favor minus motion. A high force ratio $\fM$ enhances the unbinding cascade that leads to fast motion in the plus and/or minus direction. Minus motors with high MT affinity but low force ratio tend to block motion and lead to pauses. The irregular shape of the bifurcation lines between the motility states is a discretization effect and corresponds to transitions between different locations of the maxima of the motor number probability $p$, similar to the changes of maxima locations shown in \FigMotDiagSym.

\section{Fit to the lipid-droplet data}

We applied our tug-of-war model to the bidirectional transport of lipid-droplets in \Drosophila\ embryos as studied experimentally in \mycite{WelteWieschaus98A,GrossWieschaus00A,GrossWieschaus02A}. Various transport characteristics were measured in two different embryonic phases (labeled wild type phase II and III, \WtII\ and \WtIII) and three different dynein mutation backgrounds (labeled \DhcWeak, \DhcStrong\ and \DhcBoth) during phase II.

We first considered the \WtII\ data. From cargo stall force measurements, the experimenters concluded that the droplets are pulled in the plus and minus direction by five plus and five minus motors, respectively. Since the number of active motor fluctuates stochastically, these numbers represent the average number of pulling motors. We therefore chose the total number of plus and minus motors to be $\Np=\Nm=6$. The droplets are transported by dynein \mycite{GrossWieschaus00A} and an unknown plus motor which we call kin?. The single motor parameters of dynein are only partly known, and for the stall force different labs have reported different results, see \TabMotorParameters. In the droplet experiments, cargo stall force measurements indicate a single motor stall force of 1.1 pN for both plus and minus motors\footnote{For dynein, this value is in agreement with the stall force reported by \mycite{MallikGross05A}. The low stall force for the unknown plus motor implies that this motor should be different from kinesin~1 because the kinesin~1 stall force is 6 pN.} \mycite{WelteWieschaus98A}. We used this value for the stall forces of both motors and varied the remaining 10 single motor parameters $\FdPM$, $\epsOPM$, $\piOPM$, $\vFPM$ and $\vBPM$ in order to fit the experimental data. 

We generated and analyzed cargo trajectories as described above in the section 'Numerical calculations'. In particular, we used the experimental time resolution\footnote{The spatial resolution in the experiments is of the order of nanometers and therefore unproblematic.} and definition of runs and pauses, including the experimental cutoffs. The choice of the time resolution and the cutoffs is crucial since they strongly affect the magnitude of average run lengths, velocities or pauses. For example, short runs or pauses, which are easily accessible in simulations, may be unobservable in experiments because they are below time resolution or buried in noise. 
For the experimental determination of plus and minus droplet stall forces, the droplets had to be moving in a given direction 'for a few seconds' in order to decide their direction before the measurement was performed  \mycite{WelteWieschaus98A,GrossWieschaus00A}. We therefore averaged stall force values, calculated according to \Eq{eq:cargoStallForce}, only over 'very long' runs that last more than 3 s. The experiments distinguished pauses after plus and after minus runs. We adopted this distinction although both in the experiments and in our simulations both types of pauses are very similar, see below. Furthermore, the experimenters defined 'short runs' of length 30-100 nm and 'long runs' of length 500-1000 nm, and calculated average velocities of both types of runs. We followed this procedure.

For fitting, we compared 10 transport characteristics, namely plus and minus run lengths, plus and minus stall forces, pause times after plus and minus travel, and plus and minus velocities of short and long runs\footnote{We did not use quantities that are extremely sensitive to the detectability of pauses, such as the percentage of direction reversals associated with pauses, or the average time between pauses, because pauses in simulation are more readily detected than in experiment. We also did not use 'secondary' quantities that were obtained by further processing of the data, such as fits to run length or pause time distributions or quantities calculated from these fits.} as shown in \TabFitResults. For this purpose, we defined a 'distance function' between model and experiment as the sum of squared differences between the experimentally measured and simulated quantities. As the different quantities are of different order of magnitude, they were rescaled in such a way that the experimental values are of order unity. We then minimized this distance function with respect to the unknown model parameters. 

For the \WtII\ fit, these are the 10 unknown single motor parameters listed above. We first chose 'reasonable' set of parameters. Here 'reasonable' means that the motor parameters must be of the order of magnitude of experimental single-motor parameters and that the simulation results must be of the order of magnitude of the experimental results. We then used the Nelder-Mead downhill simplex algorithm \mycite{PressFlannery02A} 
to minimize the distance function starting from this initial choice.  As this is only a local minimum, we repeated this procedure for several starting parameter sets until we found a minimum that reproduces the experimental data within about $10\%$.

In wild type phase III (\WtIII), reduced stall forces led to the conclusion that the average number of motors pulling the cargo in both directions is only four motors. We therefore took the total number of motors on the cargo to be $\Np=\Nm=5$. For the \WtIII\ fit, the single-motor parameters were set to the \WtII\ values. Then the simplex algorithm was started from this parameter set to minimize the 'distance function' with the \WtIII\ experimental values.

For the dynein mutation fits, only the six minus motor parameters $\FsM$, $\FdM$, $\epsOM$, $\piOM$, $\vFM$, $\vBM$ were used as fitting parameters because the mutation only affects the dyneins, and does so in an unknown way. The plus motor parameters were kept fixed at the values from the \WtII\ fit since droplet motion in the dynein mutation background were investigated in embryonic phase II.

The single motor parameters resulting from all these fits are shown in \TabFitPara\ (the \WtII\ values are also shown in \TabMotorParameters). They are within the expected range of motor parameters. The unbinding and binding rates are of the order of 1 s$^{-1}$ as measured for motors like kinesin~1 \mycite{ValeYanagida96A,SchnitzerBlock00A,LeducProst04A} and kinesin~3 \mycite{TomishigeVale02A}. For dynein, the unbinding and binding rates lie in the experimental range \mycite{KingSchroer00A,MallikGross05A,ReckPetersonVale06A}. The forward velocities are of the order of $0.5\,\mu$m/s which is close to the droplet velocity measured during long runs. This means that the tug-of-war does not substantially reduce the single-motor velocity. The backward velocity is two orders of magnitude smaller than the forward velocity but one order of magnitude larger than the kinesin~1 backward velocity. For dynein, this is in agreement with experiments \mycite{MallikGross05A,WangSheetz95A}. The wild type detachment forces obtained from the fit are approximately half of the stall force, similar as for kinesin~1, for which however both force scales are larger, see \TabMotorParameters. 

A comparison of the experimental data and the corresponding fit result simulation data is shown in \TabFitResults. They all agree within $10\%$. 

A remarkable feature of droplet transport is the positive correlation of run length and velocity, see \FigMotilityWtII\textit{B}: longer runs have larger velocities. The correlation persists when considering run times and velocities instead, see \FigCorrVelRunLength\textit{A}. This is more meaningful since run length and velocity are trivially linearly correlated due to the fact that high velocities lead to larger displacement.
As explained in the main text, the correlation is caused via a correlation of the average number of active winning motors with the run length, see \FigCorrVelRunLength\textit{B}. In the experiments, the correlation has been quantified by dividing the runs into short and long runs as defined above and comparing the average velocities of these runs. For visual comparison, both model and experimental results are shown in  \FigHistoCorrVelRunLength.

Although not used in the fitting procedure, the distributions of run length and of pause time show the same qualitative and similar quantitative behavior in simulation and experiment. 
\FigMotilityWtII A and \FigDoubleExp\ show the plus and minus run length distributions for all genotypes in wild type phase II, which can all be fitted by a double exponential function. The same behavior was found in the experiments \mycite{GrossWieschaus02A}. The short and long decay lengths of these fits are listed in \TabFitResultsB.  The short decay lengths are ca. 0.1 $\mu$m, while the long decay lengths are of the order of 1 $\mu$m and vary in the different genetic backgrounds. Although not used for fitting, simulation and experimental values are of the same order of magnitude and agree within 50\%.

\FigWtIIPauseTimes\ shows the pause time distribution for \WtII\ parameters. We did not distinguish pauses after plus and minus runs here because they were statistically indistinguishable. The pause time distribution can be fitted by a single exponential distribution with the time scale 0.38 s. All this is also found for the experimental distributions of pauses after plus and after minus runs, which are very similar to each other and can be fitted with a single exponential function with the time scales 0.24 s and 0.29 s, respectively \mycite{GrossWieschaus00A}. 

The agreement of simulation and experiment shows that our tug-of-war model can describe the lipid-droplet data. Two features are particularly remarkable because they are not expected within a naive picture of a tug-of-war:
First, the \WtII\ and \WtIII\ data represent motion with balanced stall forces in plus and minus direction, but the motion is net plus end-directed in phase II and net minus-end directed in phase III. 
Second, the dynein mutation data, which exhibit an impairment of both plus and minus motion, could be reproduced by varying the dynein single-motor parameters only. %
Both observations are in agreement with our tug-of-war model, as shown by the successful fit. The reason is that in our model cargo motion depends on six different motor properties (stall and detachment force, binding and unbinding rate, and forward and backward velocity) for each motor type, which leads to a rather complex behavior. In particular, cargo motion is not only determined by the motor forces but also by other motor properties, which leads to a variable response to perturbations such as mutation or regulation. As shown in \TabFitResults, it is possible (i) to change only one direction and leave the other direction unaffected (\WtIII, or change of only minus motor stall force or forward or backward velocity) (ii) to impair one direction and enhance the other (change of only minus motor unbinding or binding rate or detachment force), or (iii) to impair both directions (dynein mutations).


\end{article}

\begin{figure*}\centering
\includegraphics[scale=1.3]{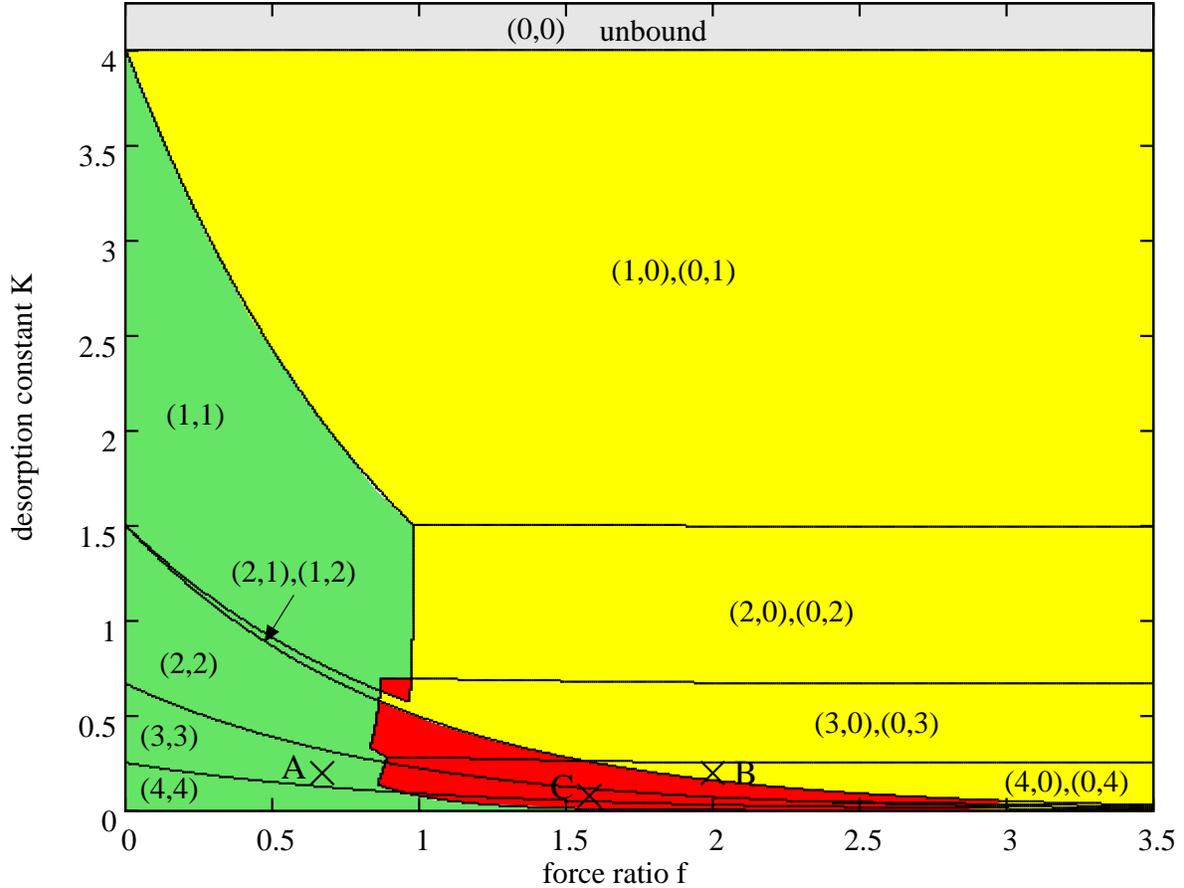}
\caption{
Motility diagram  for the tug-of-war of $\Np=\Nm=4$ symmetric plus and minus motors with identical single-motor parameters but different forward directions. Depending on the motor force ratio $f=\Fs/\Fd$ of the stall and the detachment force and the MT desorption constant $K=\epsO/\piO$, the cargo transport exhibits three different motility regimes denoted by \N, \PM, and  \PMN. These regimes are defined via the number and locations of the maxima of the motor number probability distribution $p(\np,\nm)$ as described in the text. The lines in the motility diagram separate regions in which the maxima of the probability distribution are located at different motor number states. The colors separate regions with different motility states. In motility state \N\ (green) the motor number distribution has a single maximum at a no-motion state with an equal number of plus and minus motors bound at $(\ntilde,\ntilde)$ with $1\leq\ntilde\leq 4$. The two neardiagonal maxima at (1,2) and (2,1) are also counted as a single diagonal maximum, which in a continuous state space would be at $(\ntilde,\ntilde)$ with $1<\ntilde<2$. If the maximum is at $(\ntilde,\ntilde)$=(0,0) the cargo is considered as 'unbound' (gray).
In the \PM\ regime (yellow), the probability distribution exhibits two maxima with only plus or only minus motors bound at $(\ntilde,0)$ and $(0,\ntilde)$ with $1\leq\ntilde\leq 4$.  The cargo is in the \PMN\ regime (red) if the probability distribution  exhibits three maxima. %
The parameters are kinesin-like as in \TabMotorParameters\ except for the stall force $\Fs$ and the unbinding rate $\epsO$, which are varied. The crosses labeled A, B and C mark the parameter sets for the cargo trajectories, probability and velocity distributions shown in the corresponding \FigMotilityCharSym A-C, which illustrate the qualitatively different motility behaviours for the different motility regimes. The cross B in the \PM\ regime corresponds to the complete set of kinesin parameters with $f=6/3$ and $K=1/5$, while A in the \N\ regime is at $f=2/3$ and $K=1/5$ and C in the \PMN\ regime is at $4.75/3$ and $K=0.4/5$.%
}%
\end{figure*}

\begin{figure*}\centering
\includegraphics[scale=1.2]{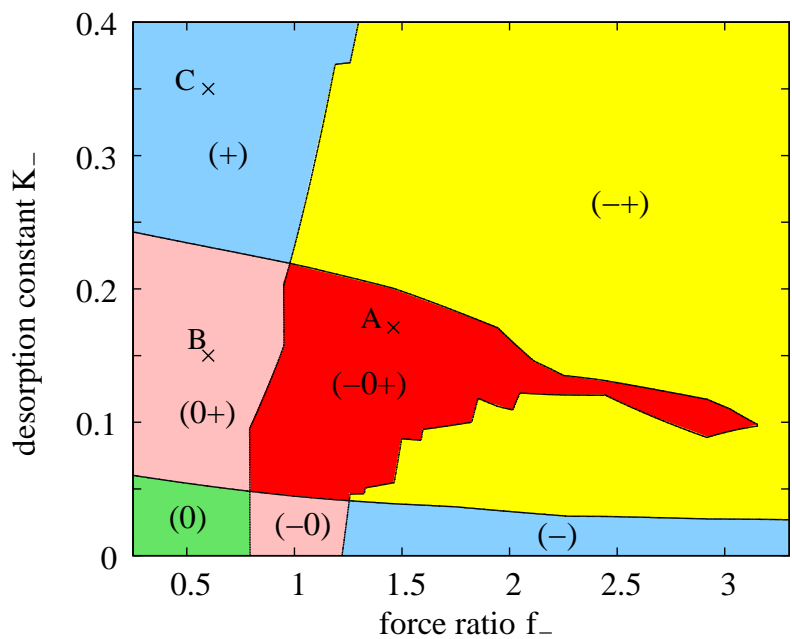}
\caption{
Motility diagram for the asymmetric tug-of-war of $\Np=6$ plus against $\Nm=6$ minus motors related to the \Drosophila\ lipid-droplet transport. 
Plus and minus motor parameters are for the \Drosophila\ plus motor (kin?) and dynein, respectively, as in \TabMotorParameters\ except for the stall force $\FsM$ and the detachment rate $\epsOM$ of the minus motors which are varied. Depending on the minus motor force ratio $\fM=\FsM/\FdM$ and the MT desorption constant $\KM=\epsOM/\piOM$, the cargo exhibits seven different motility regimes: no motion \N, fast plus motion \P, fast minus motion \M, coexistence between two of these states as indicated by \PN, \NM, and \PM,  and three-state coexistence \PMN.  The motility states are defined via the maxima of the motor number probability distribution $p(\np,\nm)$. Plus motion \P\ corresponds to a maximum at $(\nptilde,0)$ with only plus motors active, minus motion \M\ to a maximum at $(0,\nmtilde)$ with only minus motors active, and no motion \N\ to a maximum at $(\nptilde,\nmtilde)$ with both motor types active. The other motility states exhibit the combinations of these maxima as indicated by the notation. For example, in the regime \PN, the probability distribution has one maximum at a plus motion state and one maximum at a no-motion state. The crosses labeled A, B and C correspond to the parameter values for \FigMotilityCharAsym A-C, which illustrate the qualitatively distinct motility behavior for the different motility regimes. The cross A in the \PMN\ regime corresponds to the tug-of-war for \Drosophila\ droplet transport in wild type phase II with parameters as given in \TabMotorParameters. The cross B in the \PN\ regime is at $\fM=0.60$ and $\KM=0.15$, and C in the \P\ regime at $\fM=0.60$ and $\KM=0.34$.%
}%
\end{figure*}
\clearpage
\begin{figure*}[t]\centering
\includegraphics{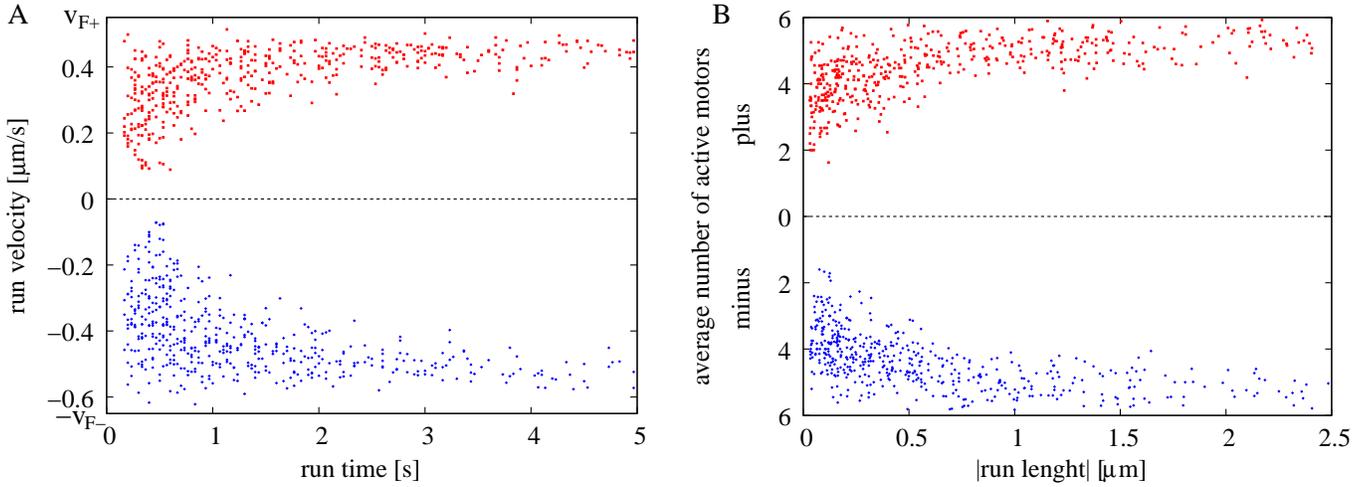}
\caption{
Scatter plots (A) of run velocities against run times and (B) of average number of active plus \resp minus motors against absolute run lengths for the \Drosophila\ droplet tug-of-war in wild type phase II for 500 runs in each direction, divided up into the positive plus runs (red) and the negative minus runs (blue). %
(A) The absolute run velocities are larger for longer runs and almost reach their maximal values of the single motor velocities $\vFP$ = 0.55 $\mu$m/s and $\vFM$ = 0.65 $\mu$m/s for very long runs. This shows that the correlation of run length and velocity discussed in the main text persists when considering run times and velocities instead. This is more meaningful since run length and velocity are trivially linearly correlated due to the fact that high velocities lead to larger displacement. %
(B) The reason for the correlation is that longer runs also have a higher average number of active pulling motors, compare the discussion in the main text. 
There are no data points for small run times, lengths and velocities because runs have been defined as having a velocity of at least 50 nm/s for at least 30 nm and 0.16 s.%
}%
\end{figure*}
\begin{figure*}[t]\centering
\includegraphics{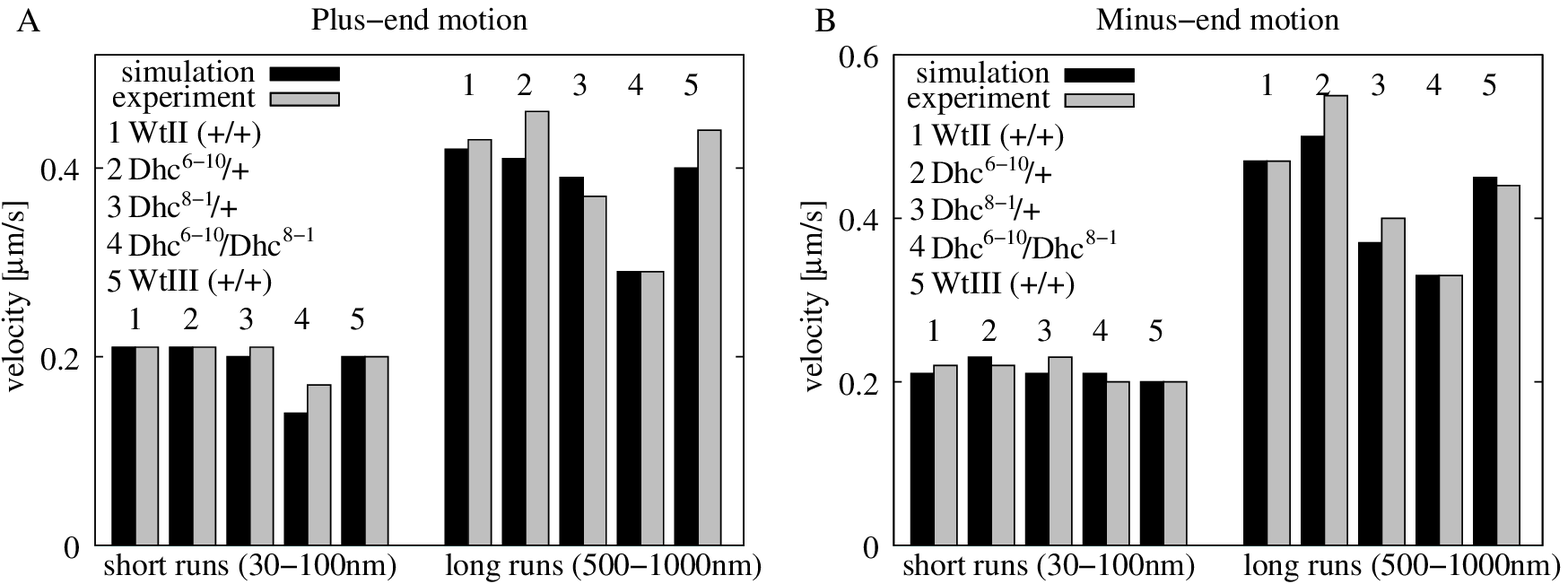}
\caption{
Mean run velocity for short (30-100 nm) and long (500-1000 nm) runs for different \Drosophila\ embryonic phases (\WtII\ and \WtIII) and dynein mutations (\DhcWeak, \DhcStrong\ and \DhcBoth) for (A) plus-end motion and (B) minus-end motion.  Short runs have shorter velocities than long runs, and changes for the different phases or genotypes are via the velocity of the long runs. The simulations values (black) are taken from \TabFitResults\ and agree well with the experimental values (gray), which are read off from Fig. 4 in Ref. \mycite{GrossWieschaus02} and Fig. 7 in Ref. \mycite{GrossWieschaus00}.%
}%
\end{figure*}
\begin{figure*}[t]\centering
\includegraphics{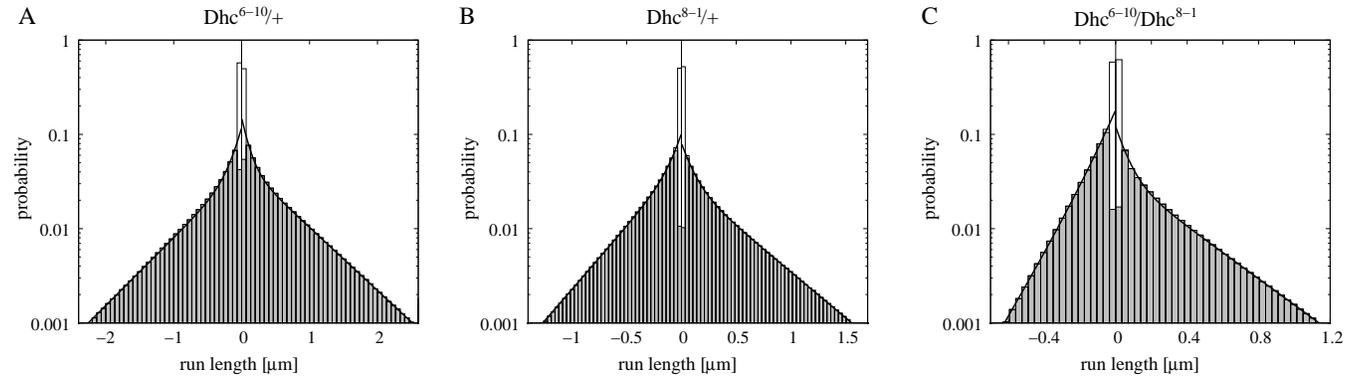}
\caption{
Run length distributions for three different dynein mutations labeled (A) \DhcWeak, (B) \DhcStrong, and (C) \DhcBoth. Plus (minus) run lengths are positive (negative). In all cases, the run length distribution can be fitted by a double exponential function corresponding to the solid line. The single motor parameters of the corresponding simulations are given in \TabFitPara. The two decay lengths of the double-exponential fits are listed in \TabFitResultsB.%
}%
\end{figure*}
\clearpage

\begin{figure*}[t]\centering
\includegraphics{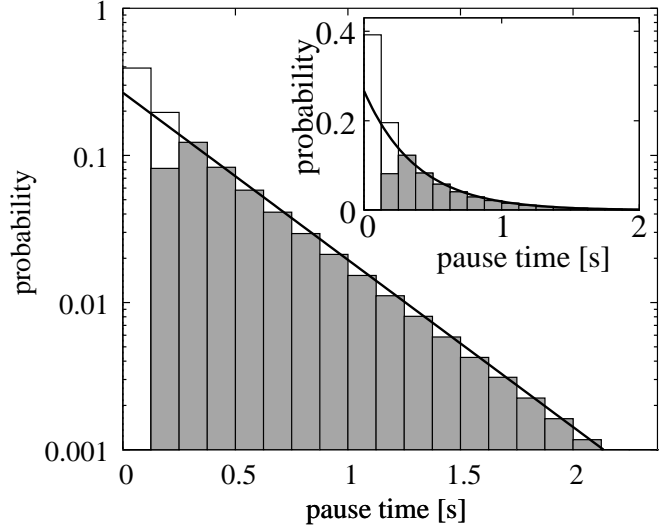}
\caption{
The pause time distribution (gray histogram) for the tug-of-war with wild type phase II parameters can be fitted by a single exponential distribution (line) with time scale 0.38 s. We did not distinguish pauses after plus and after minus motion because they are statistically indistinguishable. The full graph is a semi-logarithmic plot of the original distribution shown in the inset. The first bar has reduced counts because of the experimental definition of a pause to be longer than 0.23 s. The distribution looks similar to the experimental distributions of pauses after plus and after minus motion in Fig. 5 of Ref. \mycite{GrossWieschaus00}, which have been fitted with single-exponential distributions with time scales 0.24 s and 0.29 s. These time time scales are smaller than the experimental average pause times of 0.55 s for pauses after plus and 0.62 s for pauses after minus motion. Similarly, in simulation, the time scale 0.39 s of the exponential fit is smaller than the average pause time of 0.61 s. This indicates that the distribution is in fact not single exponential but has another, smaller time scale. This smaller time scale is below experimental resolution and shows up only when the experimental lower time cutoff 0.23 s for the pauses is removed (white histogram).%
}%
\end{figure*}

\begin{table*}[t]
\caption{
Single-motor parameters for the fits to \Drosophila\ lipid-droplet transport in wild type phase II (\WtII) and III (\WtIII), and for three different dynein mutations (\DhcWeak, \DhcStrong, and \DhcBoth). The motor numbers are $\Np=\Nm=6$ except for \WtIII\ with $\Np=\Nm=5$. The plus motor parameters for the dynein mutations are as for the plus motors in \WtII.}
\begin{minipage}{17.7cm}\small
\begin{tabular}{@{\vrule height 10.5pt depth4pt  width0pt}l||ll|l|l|l|ll}\hline
							& \multicolumn{2}{l|}{\;\;\WtII}	& \DhcWeak & \DhcStrong & \DhcBoth & \multicolumn{2}{l}{\;\;\WtIII} \\
Motor direction				& plus & minus & minus & minus & minus & plus & minus\\\hline
stall force $\Fs$ [pN]			& 1.1  & 1.1   & 1.1   & 1.1   & 0.85  & 1.1  & 1.1\\
detachment force $\Fd$ [pN]		& 0.82 & 0.75  & 0.88  & 0.84  & 1.1   & 0.82 & 0.81\\
unbinding rate $\epsO$ [s$^{-1}$]	& 0.26 & 0.27  & 0.45  & 0.37  & 0.54  & 0.26 & 0.27\\
binding rate $\piO$ [s$^{-1}$]	& 1.6  & 1.6   & 1.8   & 2.0   & 1.8   & 1.4  & 1.6\\
forward velocity $\vF$ [$\mu$m/s]	& 0.55 & 0.65  & 0.69  & 0.49  & 0.44  & 0.56 & 0.63\\
back velocity $\vB$ [nm/s]		& 67   & 72    & 77    & 76    & 53    & 68   & 73\\\hline
\end{tabular}
\end{minipage}
\end{table*}
\clearpage

\begin{table*}[t]
\caption{
Mutation and regulation in lipid-droplet transport: results of the fit to the \Drosophila\ lipid-droplet data. The first 10 lines show a comparison of the average plus and minus stall forces, the average plus and minus run lengths, the average times of pauses after plus and after minus runs, and the average velocities of short and long runs in plus and in minus direction as obtained in simulation (sim.) and experiment* (exp.) for wild type phase II  (\WtII) and III (\WtIII), and for three different dynein mutations (\DhcWeak, \DhcStrong, and \DhcBoth). The last column describes the net effect on motion (run lengths and velocities) as compared to the \WtII\ values. The last seven lines show the effect of a change of one single motor parameter from the \WtII\ value, given in front of the arrow, to the value given after the arrow in the first column.}
\begin{tabular}{@{\vrule height 10.5pt depth4pt  width0pt}l||ll|ll|ll|llll|c}\hline
		 & \multicolumn{2}{c|}{average } & \multicolumn{2}{c|}{average}  & \multicolumn{2}{c|}{average}& 
			\multicolumn{4}{c|}{average } & net\\
		 & \multicolumn{2}{c|}{\; stall force} & \multicolumn{2}{c|}{\; run length}  & \multicolumn{2}{c|}{\;pause time} 
						& \multicolumn{4}{c|}{\;velocity} & effect\\
		 & \multicolumn{2}{c|}{\;[pN]} & \multicolumn{2}{c|}{\;[$\mu$m]}  & \multicolumn{2}{c|}{\;[s]} 
						& \multicolumn{4}{c|}{\;[$\mu$m/s]} & on\\
		& & & && \multicolumn{2}{c|}{after}	& \multicolumn{2}{c}{short runs} &\multicolumn{2}{c|}{long runs} & motion\\
		 	    & plus & minus  & plus & minus      & plus & minus 	& plus & minus 	& plus & minus		&  plus/minus\\
\hline\hline
\WtII\ (sim.)	    & 5.4	& 5.3	& 0.84 & 0.66		& 0.61 & 0.61 		& 0.21 & 0.21		& 0.42 & 0.47 		& -/-\\
\WtII\ (exp.)	    & 5.5	& 5.5	& 0.84 & 0.62		& 0.55 & 0.62 		& 0.21 & 0.22		& 0.43 & 0.47 		& \\
\hline
\DhcWeak\	 (sim.) & 5.2	& 5.0	& 0.58 & 0.53		& 0.54 & 0.54 		& 0.21 & 0.23		& 0.41 & 0.50 		& impaired/impaired\\
\DhcWeak\	 (exp.) & 5.5	& 5.5	& 0.56 & 0.49		& 0.60 & 0.62 		& 0.21 & 0.22		& 0.46 & 0.55		\\\hline
\DhcStrong\ (sim.)& 5.3	& 5.1	& 0.41 & 0.32		& 0.66 & 0.66 		& 0.20 & 0.21		& 0.39 & 0.37		& impaired/impaired\\
\DhcStrong\ (exp.)& 5.1	& 5.5	& 0.38 & 0.29		& 0.71 & 0.70 		& 0.21 & 0.23		& 0.37 & 0.40		\\\hline
\DhcBoth\	 (sim.) & 5.0	& 3.9	& 0.29 & 0.15		& 0.71 & 0.75 		& 0.14 & 0.21		& 0.29 & 0.33		& impaired/impaired\\
\DhcBoth\	 (exp.) & 4.7	& 3.7	& 0.31 & 0.17		& 0.71 & 0.76 		& 0.17 & 0.20		& 0.29 & 0.33		\\\hline
\WtIII\ (sim.)    & 4.3	& 4.4	& 0.42 & 0.60		& 0.59 & 0.59 		& 0.20 & 0.20		& 0.40 & 0.45		& impaired/-\hspace*{3.5em}\\
\WtIII\ (exp.)    & 4.4	& 4.4	& 0.42 & 0.60		& -    & 0.60 		& -    & 0.20		& -	  & 0.44		\\
\hline
$\FsM=1.1\,$pN$\,\hfill\rightarrow 0.8\,$pN 			& 5.3 & 3.9  & 0.83 & 0.25  & 0.75 & 0.83  & 0.16 & 0.24  & 0.39 & 0.47 & \hspace*{3.6em}-/impaired\\ 
$\FdM=0.75\,$pN$\,\hfill\rightarrow 1.0\,$pN 			& 4.9 & 5.3  & 0.24 & 0.80  & 0.74 & 0.73  & 0.14 & 0.20  & 0.35 & 0.47 & impaired/enhanced \\
$\epsOM=0.27\,$s$^{-1}\,\hfill\rightarrow 0.5\,$s$^{-1}$& 5.5 & 4.9  & 2.0  & 0.35  & 0.44 & 0.45  & 0.27 & 0.23  & 0.45 & 0.48 & enhanced/impaired\\ 
$\piOM=1.6\,$s$^{-1}\,\hfill\rightarrow 2.5\,$s$^{-1}$ 	& 5.4 & 5.5  & 0.35 & 0.97  & 0.67 & 0.65  & 0.22 & 0.20  & 0.40 & 0.48 & impaired/enhanced\\
$\vFM=0.65\,\frac{\mu \mathrm{m}}{\mathrm{s}}\,\hfill\rightarrow 1.0\,\frac{\mu \mathrm{m}}{\mathrm{s}}$ 		
										& 5.3 & 5.4  & 0.85 & 1.4   & 0.59 & 0.60  & 0.21 & 0.24  & 0.42 & 0.70 &\hspace*{3.em}-/enhanced	\\
$\vBM=72\,\frac{\mathrm{nm}}{\mathrm{s}}\,\hfill\rightarrow 6.0\,\frac{\mathrm{nm}}{\mathrm{s}}$ 		
 										& 5.8 & 5.3  & 2.1  & 0.65  & 0.63 & 0.66  & 0.43 & 0.21  & 0.52 & 0.47 & enhanced/-\hspace*{3.2em}\\
$\Nm=6\,\hfill\rightarrow\Nm=5$ 					& 5.3 & 4.4  & 1.3  & 0.37  & 0.55 & 0.58  & 0.21 & 0.22  & 0.42 & 0.46 & enhanced/impaired\\
\hline
\end{tabular}
\tablenotes{\begin{flushleft}*The experimental values are taken from \mycite{WelteWieschaus98,GrossWieschaus00,GrossWieschaus02} as follows: The average stall forces for \WtII\ and \WtIII\ are directly given in \mycite{WelteWieschaus98}; the other stall forces are read off from the diagrams in Fig.~3 in \mycite{GrossWieschaus00} and Fig.~2, 3 in \mycite{GrossWieschaus02} by applying the procedure described in the experimental procedures of \mycite{WelteWieschaus98}. The average run lengths are from Tab.~II in \mycite{GrossWieschaus00} and Tab.~I in \mycite{GrossWieschaus02}, the average pause times from Tab.~I in \mycite{GrossWieschaus00} and Tab.~II in \mycite{GrossWieschaus02}. The average velocities for short and long runs have been read off from histograms in Fig.~7 in \mycite{GrossWieschaus00} and Fig.~4 in \mycite{GrossWieschaus02}. Missing values were not available.\end{flushleft}}
\end{table*}
\clearpage
\begin{table*}
\caption{
Mutation and regulation in lipid-droplet transport: results of the fit to the \Drosophila\ lipid-droplet data that have not been used in the fitting procedure. In all phases and genetic backgrounds, the run length distributions can be fitted by a double exponential function, see \FigMotilityWtII A and \FigDoubleExp\ with the short and long decay lengths given here. Although not used for fitting, the simulation and experimental* values are of the same order of magnitude; the maximal error is 50\%.%
}
\begin{tabular}{@{\vrule height 10.5pt depth4pt  width0pt}l||ll|ll}\hline
		 & \multicolumn{4}{c}{\; decay lengths [$\mu$m] }\\
		 & \multicolumn{2}{c|}{\; short} & \multicolumn{2}{c}{\; long} \\
		 & plus & minus & plus & minus 
\\
\hline\hline
\WtII\ (sim.)	   & 0.099	& 0.13	& 0.95	& 0.80\\
\WtII\ (exp.)	   & 0.067	& 0.098	& 1.1	& 1.1\\
\hline
\DhcWeak\	 (sim.) & 0.010	& 0.13	& 0.65	& 0.61\\
\DhcWeak\	 (exp.) & 0.088	& 0.10	& 0.78	& 0.90\\
\hline
\DhcStrong\ (sim.)& 0.094	& 0.083	& 0.45	& 0.33\\
\DhcStrong\ (exp.)& 0.074	& 0.091	& 0.40	& 0.65\\
\hline
\DhcBoth\	 (sim.) & 0.059	& 0.079	& 0.31	& 0.14\\
\DhcBoth\	 (exp.) & 0.052	& 0.044	& 0.44	& 0.21\\
\hline
\WtIII\ (sim.)    & 0.13		& 0.14	& 0.48	& 0.67\\
\WtIII\ (exp.)    & 0.096	& 0.083	& 0.78	& 1.1\\
\hline
\end{tabular}
\tablenotes{\begin{flushleft}*The experimental decay length values are from Tab.~I in \mycite{GrossWieschaus02}, and Tab.~II in \mycite{GrossWieschaus00}.\end{flushleft}}
\end{table*}

\clearpage


\begin{thebibliography}{10}

\bibitem{Gross04}
Gross SP (2004) {\em Phys Biol} 1:R1--R11.

\bibitem{Welte04}
Welte MA (2004) {\em Curr Biol} 14:R525--R537.

\bibitem{Rebhun67}
Rebhun L (1967) {\em J Gen Physiol} 50: 223--239.

\bibitem{RogersGelfand97}
Rogers SL, Tint IS, Fanapour PC,  Gelfand VI (1997) {\em Proc Natl Acad Sci USA} 94:3720--3725.

\bibitem{LigonHolzbaur04}
Ligon LA, Tokito M, Finklestein JM, Grossmann FE,  Holzbaur  ELF (2004) {\em J Biol Chem} 279:19201--19208.

\bibitem{PillingSaxton06}
Pilling AD, Horiuchi D, Lively CM,  Saxton WM (2006) {\em Mol Biol Cell} 17:2057--1068.

\bibitem{KuralSelvin05}
Kural C, Kim H, Syed S, Goshima G, Gelfand V,  Selvin PR (2005)  {\em Science} 308:1469--1472.

\bibitem{GennerichSchild06}
Gennerich A,  Schild D (2006) {\em Phys\ Biol} 3:45--53.

\bibitem{LeviGelfand06}
Levi V, Serpinskaya AS Gratton E, Gelfand VI (2006) {\em J Cell Biol.} 90: 318--327.

\bibitem{NascimentoGelfand03}
Nascimento AA, Roland JT, Gelfand VI. (2003) {\em Annu Rev Cell Dev Biol} 19: 469--491.

\bibitem{GrossGelfand02}
Gross SP, Tuma MC, Deacon SW, Serpinskaya AS, Reilein AR, Gelfand VI (2002) {\em J Cell Biol} 156:855--865.

\bibitem{WelteWieschaus98}
Welte MA, Gross SP, Postner M, Block SM, Wieschaus EF (1998)  {\em Cell} 92:547--557.

\bibitem{GrossWieschaus00}
Gross SP, Welte MA, Block SM, Wieschaus EF (2000) {\em J Cell Biol} 148:945--955.

\bibitem{GrossWieschaus02}
Gross SP, Welt, MA, Block SM, Wieschaus EF (2002) {\em J Cell Biol} 156:715--724.

\bibitem{ReileinGelfand01}
Reilein AR, Rogers SL, Tuma MC,  Gelfand VI (2001) {\em Int Rev Cytol} 204:179--238.

\bibitem{DeaconGelfand03}
Deacon SW, Serpinskaya AS, Vaughan PS, Fanarraga ML, Vernos I, Vaughan KT,  Gelfand VI (2003) {\em J Cell Biol} 160:297--301.

\bibitem{WelteGross05}
Welte MA, Cermelli S, Griner J, Viera A, Guo Y, Kim DH, Gindhart JG, Gross SP (2005) {\em Curr Biol} 15: 1266--1275.

\bibitem{SmithEnquist04}
Smith, GA, Murphy, BJ, Gross, SP,  Enquist, LW (2004) {\em Proc Natl Acad Sci USA} 101:16034--16039.

\bibitem{KaetherDotti00}
Kaether C, Skehel P,  Dotti CG (2000) {\em Mol Biol Cell} 11:1213--1224.

\bibitem{SuomalainenGreber99}
Suomalainen M, Nakano MY, Keller S, Boucke K, Stidwill RP, Greber UF (1999) {\em J Cell Biol} 144:657--672.

\bibitem{ValeBrown92}
Vale RD, Malik F,  Brown D (1992) {\em J Cell Biol} 119:1589--1596.

\bibitem{AshkinSchliwa90}
Ashkin A, Sch{\"u}tze K, Dziedzic JM, Euteneuer U,  Schliwa M (1990) {\em Nature} 348:346--348.

\bibitem{BlockerGriffiths97}
Blocker A, Severin FF, Burkhardt JK, Bingham JB, Yu H, Olivo JC, Schroer TA, Hyman AA,  Griffiths G (1997) {\em J Cell Biol} 137:113--129.

\bibitem{KlumppLipowsky05PNASCargoTransport}
Klumpp S,  Lipowsky R (2005) {\em Proc Natl Acad Sci USA} 102:17284--17289.

\bibitem{Duke00}
Duke T (2000) Phil Trans R Soc Lond B 355: 529--538.

\bibitem{GrillJulicher05}
Grill S, Kruse K, J{\"u}licher F (2005) Phys Rev Lett 94: 108104. 

\bibitem{MallikGross05}
Mallik R, Petrov D, Lex SA, King SJ,  Gross SP (2005) {\em Nature} 427:649--652.

\bibitem{WangSheetz95}
Wang Z, Khan S, Sheetz MP (1995) {\em Biophys J} 69: 2011--2023.

\bibitem{SvobodaBlock94}
Svoboda K,  Block SM (1994) {\em Cell} 77:773--784.

\bibitem{SchnitzerBlock00}
Schnitzer MJ, Visscher K,  Block SM (2000) {\em Nature Cell Biol}  2:718--723.

\bibitem{TobaHiguchi06}
Toba S, Watanabe TM, Yamaguchi-Okimoto L, Toyoshima YY,  Higuchi H (2006) {\em Proc Natl Acad Sci USA} 103:5741--5745.

\bibitem{ValeYanagida96}
Vale RD, Funatsu TS, Pierce DW, Romberg L, Harada Y,  Yanagida T (1996) {\em Nature} 380:451--453.

\bibitem{KingSchroer00}
King SJ,  Schroer TA (2000) {\em Nature Cell Biol} 2:20--24.

\bibitem{LeducProst04}
Leduc C, Campas O, Zeldovich KB, Roux A, Jolimaitre P,  Bourel-Bonnet L, Goud B, Joanny J-F, Bassereau P,  Prost J  (2004) {\em Proc Natl Acad Sci USA} 101:17096--17101.

\bibitem{ReckPetersonVale06}
Reck-Peterson SL, Yildiz A, Carter AP, Gennerich A, Zhang N, Vale RD (2006) {\em Cell} 126:335--348.

\bibitem{CarterCross05}
Carter NJ,  Cross RA (2005) {\em Nature} 435:308--312.

\bibitem{NishiuraSutoh04}
Nishiura M, Kon T, Shiroguchi K, Ohkura R, Shima T, Toyoshima YY, Sutoh K (2004) {\em J Biol\ Chem} 22:22799--22802.

\end{thebibliography}

\begin{thebibliography}{10}

\bibitem{SvobodaBlock94A}
Svoboda K, Block SM (1994) {\em Cell} 77: 773--784.

\bibitem{KojimaYanagida97A}
Kojima H, Muto E, Higuchi H, Yanagida T (1997) {\em Biophys J} 73: 2012--2022.

\bibitem{VisscherBlock99A}
Visscher K, Schnitzer MJ, Block, SM (1999) {\em Nature} 400: 184--189.

\bibitem{TobaHiguchi06A}
Toba S, Watanabe TM, Yamaguchi-Okimoto L, Toyoshima YY,  Higuchi H (2006) {\em Proc Natl Acad Sci USA} 103: 5741--5745.

\bibitem{CarterCross05A}
Carter NJ, Cross RA (2005) {\em Nature} 435: 308--312.

\bibitem{KlumppLipowsky05PNASCargoTransportA}
Klumpp S, Lipowsky R (2005) {\em Proc Natl Acad Sci USA} 102: 17284--17289.

\bibitem{MallikGross04RevA}
Mallik R, Gross SP (2004) {\em Curr Biol} 14: R971--R982.

\bibitem{WelteWieschaus98A}
Welte MA, Gross SP, Postner M, Block, SM,  Wieschaus EF (1998) {\em Cell} 92: 547--557.

\bibitem{GrossWieschaus00A}
Gross SP, Welte MA, Block SM,  Wieschaus EF (2000) {\em J Cell Biol} 148: 945--955.

\bibitem{GrossWieschaus02A}
Gross SP, Welte MA, Block SM,  Wieschaus EF (2002) {\em J Cell Biol} 156: 715--724.

\bibitem{vanKampen92A}
van Kampen NG (1992) {\em Stochastic processes in physics and chemistry} (Elsevier, Amsterdam).

\bibitem{Gillespie76A}
Gillespie DT (1976) {\em J Comp Phys} 22: 403--434.

\bibitem{MallikGross05A}
Mallik R, Petrov D, Lex SA, King SJ,  Gross SP (2005) {\em Nature} 427: 649--652.

\bibitem{PressFlannery02A}
Press WH, Teukolsky SA, Vetterling WT,  Flannery BP (2002) {\em Numerical recipes in C++ : the art of scientific computing} (Cambridge: Cambridge University Press, New York).

\bibitem{ValeYanagida96A}
Vale RD, Funatsu TS, Pierce DW, Romberg L, Harada Y,  Yanagida T (1996) {\em Nature} 380: 451--453.

\bibitem{SchnitzerBlock00A}
Schnitzer MJ, Visscher K,  Block SM (2000) {\em Nature Cell Biol} 2: 718--723.

\bibitem{LeducProst04A}
Leduc C, Campas O, Zeldovich KB, Roux A, Jolimaitre P,  Bourel-Bonnet L, Goud B, Joanny J-F, Bassereau P,  Prost J  (2004) {\em Proc Natl Acad Sci USA} 101:17096--17101.

\bibitem{TomishigeVale02A}
Tomishige M, Vale RD (2002) {\em Science} 297: 2263--2267.

\bibitem{KingSchroer00A}
King SJ, Schroer TA (2000) {\em Nature Cell Biol} 2: 20--24.

\bibitem{ReckPetersonVale06A}
Reck-Peterson SL, Yildiz A, Carter AP, Gennerich A, Zhang N, Vale RD (2006) {\em Cell} 126:335--348.

\bibitem{WangSheetz95A}
Wang Z, Khan S, Sheetz MP (1995) {\em Biophys J} 69: 2011--2023.


\end{thebibliography}
\end{document}